\documentclass{ws-rv9x6}  
\usepackage{amssymb}
\usepackage{graphicx}

\def\0{\over } \def\2{{1\over2}} \def\4{{1\over4}}
\def\5{\hat } \def\6{\partial }

\def\a{\alpha } \def\b{\beta }  \def\d{\delta }

\def\nbfgrad{\mbox{\boldmath$\nabla$}}

\def\({\left(} \def\){\right)} \def\<{\langle } \def\>{\rangle }

\def\Tr{{\,\rm Tr\,}}
\def\Im{{\,\rm Im\,}}
\def\Re{{\,\rm Re\,}}
\def\tr{{\,\rm tr\,}}

\newcommand{\be}{\begin{equation}}
\newcommand{\ee}{\end{equation}}
\newcommand \beq{\begin{eqnarray}}
\newcommand \eeq{\end{eqnarray}}
\newcommand{\bea}{\begin{eqnarray}}
\newcommand{\eea}{\end{eqnarray}}

\newcommand{\nn}{\nonumber\\ }

\begin{document}

\setcounter{chapter}{0}

\chapter*{
THERMODYNAMICS OF THE\\
HIGH-TEMPERATURE QUARK--GLUON PLASMA
}
\markboth{J.-P. Blaizot, E. Iancu, A. Rebhan}{Thermodynamics of the
High-Temperature Quark--Gluon Plasma}
\author{%
Jean-Paul Blaizot 
 and
Edmond Iancu
}
\address{Service de
Physique Th\'eorique, CEA/DSM-Saclay,\\
Unit\'e de Recherche Associ\'ee au CNRS,
91191 Gif-sur-Yvette, France}
\author{Anton Rebhan
}
\address{Institut f\"ur Theoretische Physik, Technische
Universit\"at Wien,\\
Wiedner Hauptstr. 8-10, 
A-1040 Vienna, Austria}


\vskip 1cm
\begin{abstract}
We review the various methods which have been employed
recently to describe the  thermodynamics of the high
temperature quark-gluon plasma using weak coupling
techniques, and we compare their results with those
of most recent  lattice gauge calculations. Many of
the  difficulties encountered with
perturbation theory at finite temperature
are in fact not specific to QCD but are  present
in any field theory at finite temperature and will be
discussed first in the simple example of the scalar field theory.
We discuss the merits and limitations
of various techniques which have been used  to  go beyond
perturbation theory in the soft sector, such as dimensional reduction,
screened perturbation theory or hard-thermal-loop perturbation theory,
and  $\Phi-$derivable approximations. In the last part of the
review, we focus on the later, which  lead to a remarkably simple
expression for the entropy of the quark-gluon plasma. When complemented
with further, physically motivated, approximations, this approach
reproduces accurately  the entropy obtained from lattice gauge
calculations  at temperatures above
$2.5 T_c$, where $T_c$ is the deconfinement temperature. This
calculation thus provides also  support to the physical picture of the
quark-gluon plasma as a gas of weakly interacting quasiparticles.
\end{abstract}

\newpage
\tableofcontents
\newpage

\section{Introduction}

Much of the motivation for studying heavy-ion collisions at 
ultra-relativistic energies is based on
the expectation that matter at high temperature and/or density 
becomes simple: because of the
asymptotic freedom of Quantum Chromodynamics (QCD), one expects 
that in this regime matter becomes a
weakly interacting gas of quarks and gluons whose properties should 
be calculable in a weak coupling
expansion.

The existence of weakly interacting quark matter was indeed
anticipated on the
basis of asymptotic freedom of QCD\cite{Collins:1975ky}. But the most
compelling theoretical evidences for the existence of  the
quark-gluon plasma are coming from lattice gauge calculations (for recent
reviews see e.g. Refs.%
\cite{Karsch:2001cy,Laermann:2003cv}). These are at present the
unique tools allowing  a detailed study of the transition region where
various interesting phenomena are taking place, 
such as colour deconfinement or
chiral symmetry restoration. In this review, we shall not consider this
transition region,\footnote{For
new ideas on how to cover this region by effective field theory
methods see Refs.~\cite{Pisarski:2000eq
}.},
but focus rather on the high temperature phase, 
exploiting
the fact that at sufficiently high temperature 
the effective gauge coupling
constant should be small enough to allow for weak coupling calculations%
\cite{Kap:FTFT,LeB:TFT,Smilga:1996cm,Blaizot:2001nr}.

The physical picture which emerges from these
weak coupling calculations is a simple one, 
and in many respect the quark-gluon
plasma is very much like  an ordinary  electromagnetic
plasma  in the ultrarelativistic regime (with however specific
effects related to the nonabelian gauge symmetry%
\cite{Braaten:1990mz,Frenkel:1990br,Blaizot:1995ws}). 
To zeroth order in an expansion
in powers of the coupling $g$, the quark gluon plasma is a gas of
noninteracting
quarks and gluons, with typical momenta $k\sim T$. This is the ideal 
quark-gluon plasma. The interactions
appear to alter only slightly this simple physical picture: they 
turn   plasma particles  into
\textit{massive quasiparticles}, and generate \textit{collective 
modes} at small momenta. However, while
the relevant degrees of freedom have been identified long ago, and 
effective theories to describe their
dynamics are known%
\cite{Braaten:1990mz,Blaizot:1993gn,
Blaizot:1994be,%
Nair:1993rx,
Blaizot:1999xk}, 
the calculation of the thermodynamics
of the quark-gluon plasma using weak coupling techniques has remained 
unsuccessful until recently.

Strict expansions in power of the coupling constant  have been pushed 
up to order $g^5$,  both in QCD
and in scalar field theories. In both cases, one observes a rather 
poor apparent convergence,
the successive contributions oscillating wildly, unless the coupling 
is very small and the deviation from ideal-gas behaviour
correspondingly minute. As we shall see, the
difficulty  may be related to  the fact that what determines the 
accuracy of a weak coupling calculation in
thermal field theories is not only the strength of the coupling, but 
also the magnitude of the thermal
fluctuations. These vary according to the relevant momentum scales, 
so that the accuracy of the weak coupling
expansion depends on which momentum scale contributes dominantly to 
the quantity under consideration.

To get
orientation into these effects, let us then consider a massless 
scalar field theory with
$g^2\phi^4$ interactions.
In the noninteracting case  the magnitude of the thermal 
fluctuations of the field is given by the simple
formula (with
$\varepsilon_k=k$):
\beq\label{fluctuationsA}
\langle \phi^2\rangle=
\int \frac{{\rm d}^3
k}{(2\pi)^3}\frac{n(\varepsilon_k)}{\varepsilon_k},
\eeq
where $n(\varepsilon_k)=1/({\rm e}^{\beta \varepsilon_k}-1)$ is the 
Bose-Einstein distribution
function.  We shall use this formula as an approximation for the 
magnitude of the fluctuations also  in
the interacting case, eventually modifying $\varepsilon_k$
appropriately.
     In order to get a criterion for the expected
validity of perturbation theory, we shall compare the expectation 
value of the ``kinetic
energy'' $\langle( \partial \phi)^2\rangle$ with the ``interaction 
energy'' which we approximate as
$g^2\langle\phi^2\rangle^2$.

    For
the plasma particles   $\varepsilon_k=k\sim T$ and  $\langle
\phi^2\rangle_T\sim T^2$. Then indeed, when
$g$ is small,   the kinetic energy  of a typical
particle, $k^2\sim T^2$, is large compared to its interaction energy 
$\sim g^2\langle\phi^2\rangle_T$:
The  short wavelength, or {\it hard},  fluctuations produce a small 
perturbation on the motion of a plasma
particle. However, this is not so for an excitation at the momentum 
scale  $k\sim gT$:  then  the kinetic
energy
$k^2\sim (gT)^2$ is comparable to the contribution to the interaction
energy coming from the coupling to the hard modes, 
$g^2\langle\phi^2\rangle^2_T$. Thus, the
properties of an excitation with momentum
$gT$ are expected to be nonperturbatively renormalized
by the hard thermal fluctuations. And indeed, the scale
$gT$ is that  at which collective phenomena develop. The emergence of 
the Debye screening mass $m_D\sim gT$
in a plasma is one of the simplest examples of such phenomena. More 
generally, this renormalization of the
soft mode dynamics by hard mode contributions invites a description in terms of
an effective theory for the soft modes, the parameters of this 
effective theory being determined by the hard
modes. The building blocks of such an effective theory have  been 
dubbed \textit{hard thermal loops} (HTL). In
the scalar field theory, there is only one such HTL, which is a mass. 
But in QCD, there is an infinite number
of them\cite{Braaten:1990mz,Frenkel:1990br}.

Let us now consider the fluctuations at the {\it soft} scale
$gT\ll T$. These fluctuations can be accurately described by classical
fields. In fact the associated occupation numbers
$n(\varepsilon_k)$ are large, and
accordingly one can replace
$n(\varepsilon_k)$ by $ T/\varepsilon_k$ in eq.~(\ref{fluctuationsA}).
Introducing  an upper cut-off $gT$ in the momentum integral, one  then
gets:
\beq
\langle \phi^2\rangle_{gT} \sim \int^{gT}{\rm d}^3k \, \frac{T}{k^2}\sim
gT^2.
\eeq
Thus $g^2\langle\phi^2\rangle_{gT}
\sim g^{3}T^2$ is still of higher order than the kinetic term $g^2T^2$.
In that sense the soft modes with $k\sim gT$ are still perturbative. 
Note however that
they generate contributions to physical observables which are not analytic in
$g^2$, as shown by the example of the order
$g^3$ contribution to the
energy density, which goes schematically as follows:
\beq
\epsilon^{(3)}\sim  \int_0^{gT}
{\rm d}^3 k \,\,\varepsilon_k\,\frac{T}{\varepsilon_k}\sim
T(gT)^3\,\sim\, g^3T^4.
\eeq
In fact, in
contrast to the perturbation theory for the hard modes, the parameter 
which controls the
perturbative expansion for the soft modes is $g$ instead of $g^2$ so 
that is it less accurate in general. 
As we shall see, much of the difficulties with 
the strict perturbative expansion can
be attributed to the poor accuracy of perturbation theory in the soft 
sector, and the present review
describes various solutions proposed to get around this problem.

The previous arguments concerning the fluctuations can be extended to 
QCD, where one should compare  the
kinetic energy
$(\partial A)^2$ with the magnitude of the nonabelian interactions $g^2 
A^4$ (with $A$ the gauge potential). For
the plasma particles   $\varepsilon_k=k\sim T$ and  $\langle
A^2\rangle_T\sim T^2$.
The associated electric (or magnetic) field fluctuations are
$\langle E^2 \rangle_T\sim
\langle (\partial A)^2\rangle_T \sim k^2 \langle A^2\rangle_T\sim
T^4
$ and are a dominant contribution to the plasma energy density.
    As already mentioned, these short wavelength,
or {\it hard}, gauge field
fluctuations produce a small perturbation on the motion of a plasma particle.
However, this is not so for an excitation at the momentum scale 
$k\sim gT$ which is  nonperturbatively
renormalized
by the hard thermal fluctuations. And again perturbation theory for 
the soft modes is governed by
$g$, not $g^2$.

A new feature emerges however in QCD:   at the  {\it ultrasoft } 
momentum scale, $k\sim g^2T$,  the
unscreened magnetic fluctuations  play a dominant role.  At that 
scale  it becomes necessary to distinguish
the electric and the magnetic sectors (which provide comparable 
contributions at the scale
$gT$). The  electric
fluctuations decouple
because the Debye screening mass $m_D^2\gg k^2\sim (g^2T)^2$
and their contribution is  negligible, of order  $ g^4
T^2$. However, because of
the absence of static screening in the magnetic sector, we have there
$\varepsilon_k\sim k$ and
\beq\label{fluctg2t}
\langle A^2\rangle_{g^2T}\sim T\int_0^{g^2T}{\rm d}^3k \frac{1}{k^2}\,
\sim\,g^2 T^2,\eeq
so that $(\partial A)^2\sim g^6T^4\sim g^2 (A^2)^2$: the fluctuations are 
no longer perturbative. This
is the origin of the well-known breakdown of perturbation theory at 
order $g^6$.

Whereas  one may argue that these non-perturbative magnetic fluctuations can
contribute to the deviation of the pressure from the ideal gas result, as
observed on the lattice, this contribution, being of order $g^6$, is 
presumably numerically rather small sufficiently above the transition
temperature.
At any rate, it is clear that the magnetic fluctuations have a priori no 
responsibility for the lack of convergence of
the weak coupling expansion at lower orders (i.e., up to order
$g^5$). This is further confirmed by
the fact that a similar lack of convergence is seen also in other theories,
like the scalar field theory. These magnetic fluctuations require 
genuine non-perturbative treatment, such as
that based on a combination of dimensional reduction and lattice 
calculations\cite{Kajantie:2000iz}. This will be
briefly discussed. But most of the review will focus on purely analytical
resummation techniques which have been
proposed to improve perturbation theory in the soft sector.

Among those, the simplest one which attempts to account non%
perturbatively for the dominant effect of
interactions in the soft sector, which in the case of the scalar 
field is to generate a mass for the
excitations, is the so-called   ``screened perturbation theory''%
\cite{Karsch:1997gj,Chiku:1998kd,Andersen:2000yj}. 
This consists in a reorganization of the 
perturbative expansion, based on the following
rewriting of the Lagrangian:
\be
{\cal L}= {\cal L}_0-\frac{1}{2}m^2\phi^2+
\frac{1}{2}m^2\phi^2+{\cal L}_{int}
= {\cal L}_0^\prime +{\cal L}_{int}^\prime \, ,
\ee
with ${\cal L}_0^\prime={\cal L}_0-(1/2)m^2\phi^2$.
  A perturbative expansion in
terms of screened propagators (that is keeping the screening mass $m$ as
a parameter, i.e. not as a perturbative correction to be expanded out)
has been shown to be quite stable  with good
convergence properties. However,
$m$ depends on the temperature, which makes the propagators 
explicitly temperature dependent, and  the
ultraviolet divergences which occur in high order calculations become 
temperature
dependent.  
Thus, at any finite loop order, the ultraviolet renormalization
gets artificially modified.

In the case of gauge theory, the effect of the interactions
is more complicated than just generating a mass. But we know how to determine
the dominant corrections to the self-energies. When the momenta are soft, these
are given by the hard thermal loops\cite{Braaten:1990mz,Frenkel:1990br}. 
By adding these
corrections to the tree level Lagrangian, and subtracting them from the
interaction part, one generates the so-called hard thermal loop perturbation
theory (HTLPT)\cite{Andersen:1999fw
}:
\beq
{\cal L}={\cal L}_0+{\cal L}_{HTL}- {\cal L}_{HTL}+{\cal L}_{int}
= {\cal L}_0^\prime +{\cal L}_{int}^\prime\, .
\eeq
  The resulting
perturbative expansion is made complicated however by the nonlocal 
nature of the hard
thermal loop action, and by the necessity of introducing temperature 
dependent counter
terms.  Also, in such a scheme, one is led to use the hard thermal 
loop approximation in
kinematical regimes where it is no longer a justifiable approximation.

$\Phi$-derivable approximations\cite{Baym:1962} 
provide a more natural 
approach to propagator renormalization
than screened perturbation theory, although the general goal is 
similar. It allows us to exploit fully
the spectral information that one has about the plasma
excitations. In particular the hard thermal loops enter as essential 
ingredients,\cite{Blaizot:1999ip,Blaizot:1999ap,Blaizot:2000fc} 
but in contrast to HTLPT,
they contribute in an essential way only for
external momenta where they are accurate. 
One of the main advantages of such
approximations is that they lead to remarkable simplifications in the 
calculation of the entropy.\cite{Vanderheyden:1998ph} This
allows one, in particular, to bypass some of  the difficulties of 
HTLPT. Besides, the entropy makes transparent the underlying 
physical picture of quasiparticles, which (in simpler models) has
been shown to provide rather accurate  fits to lattice data%
\cite{Peshier:1996ty,Levai:1997yx,Peshier:1999ww,Schneider:2001nf}. 
The  quasiparticle picture assumes that
the dominant effect  of the interactions can be incorporated in the 
spectral properties of suitably defined
quasiparticles with small residual interactions; this is precisely 
what comes out of the entropy calculation.

Finally, we should mention other attempts to improve the behaviour 
of perturbation theory by using various
mathematical extrapolation schemes.
Thus, extrapolations have been constructed, based on  the 
first terms of the perturbative series, using Pad\'e
approximants\cite{Kastening:1997rg,Hatsuda:1997wf,Cvetic:2002ju}
or Borel summation techniques\cite{Parwani:2000rr
}. 
The resulting expressions are
indeed  smooth functions of the coupling, better behaved than polynomial
approximations truncated at order $g^5$ or lower, with a weak dependence on the
renormalization scale.  However these techniques, which offer
little physical insight, will not be discussed further here.

The outline of the paper is as follows. In the next section, we shall 
review various features of perturbation
theory in quantum field theory at finite temperature, using the 
simple example of the scalar field. This will
offer us the possibility to illustrate most of the difficulties 
encountered in thermal gauge theories,
  and also
the various resummation schemes which have been proposed in the 
literature. Section 3 presents a brief 
summary of the state of the art concerning the perturbative 
calculations in QCD,  those based on
dimensional reduction, as well as recent lattice data. In section 4 
we describe  the calculation of the
entropy and emphasize its technical simplicity and its physical 
content. The last section of the review
summarizes the various conclusions and puts the different approaches 
into a general perspective.

\section{The scalar field theory as a 
pedagogical example}

As mentioned in the introduction, many of the difficulties of weak 
coupling calculations at high
temperature are not specific to gauge theories. They will be 
illustrated in this section by studying
the thermodynamics of a scalar field  with Lagrangian
\beq\label{Lagran}
{\cal L}&=& {1\over2}\partial_\mu\phi\partial^\mu\phi-
{m^2\over2}\phi^2 - {\lambda\over 4!}\phi^4\,.\eeq
We shall  often use  $g^2 \equiv\lambda/4! $ as an alternative 
notation for the coupling
strength.

\subsection{Perturbation theory and its difficulties}

The weak--coupling expansion of the pressure in the case $m=0$
has been computed to order $\lambda^{5/2}$ (or $g^5$),
and reads\cite{Arnold:1994ps,Parwani:1995zz,Braaten:1995cm}:
\beq\label{Pphi}\nonumber
P &=& P_0\Bigg[
1-{15\over 8}\left(\frac{g}{\pi}\right)^2
+{15\over 2}\left(\frac{g}{\pi}\right)^3 +{135\over 16}
\left(\log{\bar\mu\over2\pi T}+0.4046\right)\left(\frac{g}{\pi}\right)^4
\\
&{}&\quad -{405 \over 8}\left(\log{\bar\mu\over2\pi
T}-{4\over3}\log\frac{g}{\pi}
-0.9908 \right)\left(\frac{g}{\pi}\right)^5
+{\cal O}(g^6 \log g)\Bigg]\;,
\eeq
where $P_0 = (\pi^2/90)T^4$
is the pressure of an ideal gas of free massless bosons,
and $ g^2(\bar\mu) \equiv \lambda(\bar\mu)/24$ is the
$\overline{\rm MS}$ coupling constant at the renormalization scale $\bar\mu$.
This formula calls for several remarks.

First we note that,
to order $\lambda^{5/2}$, $P$ is \emph{formally } independent of $\bar\mu$
(in the sense that $dP/d\bar\mu$ involves terms of order
   $\lambda^3$ at least). This is easily verified using the
renormalization group equation satisfied by the renormalized coupling
$\lambda(\bar\mu)$:
\beq\label{renorm0}
\bar\mu \frac{{d}\lambda}{d\bar\mu } =3\,
\frac{\lambda^2}{16\pi ^{2}}+{\mathcal{O}}\left( \lambda^3\right) .
\eeq
But the approximate expression (\ref{Pphi}) for $P$ {\it does} depend
numerically on $\bar\mu$. In order to avoid large logarithms
$\sim \ln({\bar\mu }/{2\pi T})$ in the expansion (\ref{Pphi}) one may
choose $\bar\mu \simeq 2\pi T.$ With this choice
$\lambda\left( \bar\mu
\right) $ becomes effectively a function of the temperature. In the
case of QCD where the $\beta$-function is negative, this leads to the
expectation that the effective coupling becomes small at large
temperature.

\begin{figure}[tb]
\centerline{\includegraphics[bb=70 200 540 540,width=6.5cm]{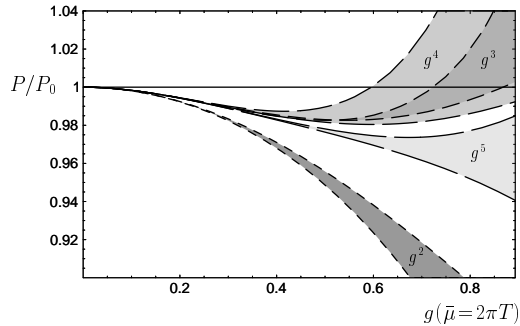}}
\caption{Weak-coupling expansion through
orders $g^2$, $g^3$, $g^4$, and $g^5$
for the pressure normalized to that of an ideal gas as a function
of $g(2\pi T)$ in $\phi^4$ theory.}
\label{fpert}
\end{figure}

In Fig.~\ref{fpert}, we show the successive perturbative approximations to
$P/P_0$ as a function of $g(2\pi T)$. Each partial
sum is shown as a band obtained by varying $\bar\mu$
from $\pi T$ to $4\pi T$, which gives an indication of the
theoretical uncertainty in the calculation. Thus, in drawing this plot,
$g(\bar\mu)$ has been calculated from
$g(2\pi T)$ by using the renormalization group equation (\ref{renorm0}).
The lack of convergence of the weak-coupling expansion
is evident in Fig.~\ref{fpert}.
We can infer from Fig.~\ref{fpert} that the expansion in powers of $g$
ceases to make sense as soon as $g(2 \pi T)\gtrsim 0.4$.

Another remark concerns the presence of terms non--analytic in
the coupling constant $\lambda$, i.e., terms of order $g^3\sim \lambda^{3/2}$,
$g^5\sim \lambda^{5/2}$, and also terms involving $\ln g$, in the perturbative
expansion in eq.~(\ref{Pphi}). As we shall see shortly, these terms
result from infinite resummations of subsets of diagrams of perturbation
theory.

At this stage, it is instructive to compare with the corresponding expansion
of the  {\it screening} ``Debye'' mass $m_D$, defined from the
the static propagator by\footnote{A definition
through the location of the pole rather than the infrared limit
of the self-energy is crucial to make $m_D$ gauge-independent
in non-Abelian theories%
\cite{Kobes:1990xf
};
this definition is also renormalization-group
invariant\cite{Rebhan:1993az,
Blaizot:1995kg}.}
\bea
\label{scrdef}
{\bf p}^2 + m^2_D
+\Pi(0,{\bf p}) = 0 \hspace{1cm}
          {\rm at} \;\;\; {\bf p}^2=-m_D^2 \;,
\eeq
where $\Pi(p_0,{\bf p})$ is the self--energy.
For the massless case ($m=0$), the weak coupling expansion of
$m_D^2$ is known to order $g^4$, and reads\cite{Braaten:1995cm}:
\beq
\label{mpert}
\!\!m_D^2=g^2T^2\Big\{1- 3 \frac{g}{\pi}
-\frac{9}{2}
\left[  \ln{\bar\mu\over2\pi T}- {4\0 3}\ln {g\0 \pi}-2.415 \right]
\!\left(\frac{g}{\pi}\right)^2+{\cal O}(g^3)\Big\}.
\eeq
Note that, with our notation $g^2\equiv \lambda/24$ for the
coupling constant, the leading--order contribution in eq.~(\ref{mpert}) is
simply
\beq
\label{mhat}
\hat m_D\,=\,gT\,,\eeq
which happens to be the same relation as in pure-glue QCD
to leading order, cf.~(\ref{MD}).
As for the pressure, non--analytic terms ($g^3$, $\ln 1/g$)
occur in the expression of $m_D^2$.

\begin{figure}[tb]
\centerline{\includegraphics[bb=70 200 540 540,width=6.5cm]{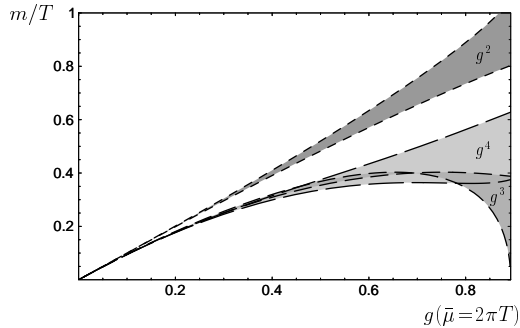}}
\caption{Weak-coupling expansion to orders $g^2$, $g^3$, and $g^4$
for the screening mass normalized to the temperature as a
function of $g(2\pi T)$.}
\label{mspert}
\end{figure}

In Fig.~\ref{mspert}, we show the screening mass $m_D$
normalized to the leading order result $\hat m_D$
as a function of $g(2\pi T)$, for each of the three
successive approximations to $m_D^2$. As in Fig.~\ref{fpert},
the bands correspond to varying $\bar\mu$
from $\pi T$ to $4\pi T$. The poor convergence is again evident, with a pattern
  similar to that in Fig.~\ref{fpert}: a large difference between the 
order-$g^2$ and order-$g^3$
approximations, and a larger sensitivity to the value of the 
renormalization scale for the order
$g^4$.

In order to get some insight on this peculiar behaviour of the
expansions (\ref{Pphi}) and (\ref{mpert}), we shall briefly recall
how the various terms are obtained, up to order $g^3$.

We
start with the zeroth
order calculation. We have $P_0=-\Omega_0/V$, with
\beq\label{O0}
\frac{\Omega_0}{V}
&=&\2\,T\sum_{n}
   \int\frac{{\rm d}^3k}{(2\pi)^3}\,\ln (\omega_n^2 + {\bf k}^2 + m^2)
\nn
&=&\2
\int\! {d^3k\0(2\pi)^3}\,\varepsilon_k\,\,+\,\,
T\int\! {d^3k\0(2\pi)^3}\,\log(1-{\rm
e}^{-\varepsilon_k/T}),
\eeq
where
$\omega_n=2n\pi T$
is a bosonic Matsubara frequency and $\varepsilon_k^2=k^2+ m^2$.
The first term in the second line of eq.~(\ref{O0})
is the sum of the `zero-point' energies of free massive bosons. This 
is ultraviolet
divergent, but independent of the temperature; it can therefore be 
absorbed in the
redefinition of the vacuum energy, and  discarded.
The second term
is temperature--dependent and finite; this is recognized as minus the
pressure of the ideal gas of scalar particles. For  $m\ll T$ this
can be expanded as:
\beq\label{P0mexp}
P_0(m)=-T\int\frac{d^3k}{(2\pi)^3}\ln
(1-e^{-\beta\varepsilon_k})
= \frac{\pi^2T^4}{90} -\frac{m^2 T^2}{24}+
\frac{ m^3 T}{12\pi} +\ldots ,\;
\eeq
where the neglected terms start at order $m^4\ln(m/T)$.

\begin{figure}[b]
\epsfxsize=2cm
\centerline{\epsffile{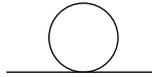}}
\vspace{3mm}
\caption[a]{Lowest order correction to the self-energy in scalar
$\phi^4$ theory.}
\label{fig:tadpole}
\end{figure}

Consider now the contributions of order $g^2$
  to the self--energy and the thermodynamic potential, denoted
as $\Pi_2$ and $\Omega_2$, respectively. $\Pi_2$ is
given by the ``tadpole'' diagram of Fig.~\ref{fig:tadpole}.
This gives:
\beq\label{Sigma}
\Pi_2&=&
{\lambda\over 2}\,T\sum_{n}
   \int\frac{{\rm d}^3k}{(2\pi)^3}\,\frac{1}{\omega_n^2 + {\bf k}^2 + 
m^2}\equiv {\lambda\over 2}\,
I(m)\nn &=&{\lambda\over 2}\int\!{d^3k\0(2\pi)^3}\,\frac{1+2n(\varepsilon_k)}
{2\varepsilon_k}\equiv {\lambda\over 2}\left[ I_0(m)+I_T(m)\right]\,,
\eeq
where $n(\omega)$ is the
Bose-Einstein thermal distribution:
\beq
n(\omega)\,=\,\frac{1}{{\rm e}^{\beta \omega}\,-\,1}\,,\eeq
and the notation $I(m)$ for the sum-integral, as well as $I_0(m)$ and
$I_T(m)$ for its zero temperature and thermal contributions, will be 
used repeatedly in the
following.  The first term in the second line of eq.~(\ref{Sigma}) is 
ultraviolet divergent, but
independent of the temperature, so it can be absorbed into the 
definition of the
(zero--temperature) mass. For instance, is we choose the mass 
parameter $m$ in the Lagrangian
to be the physical mass at $T=0$, we must add a counterterm $\2
\delta m^2\phi^2$  in the Lagrangian, with
$\delta m^2$ chosen so that: 
\beq\label{counterterm}
{\lambda\over 2}\,I_0(m)+\delta m^2=0.
\eeq
Then, the lowest--order self--energy
  correction due to the thermal fluctuations
reads:
\beq\label{Sigma2}
\Pi_2={\lambda\over 2}\,I_T(m)\approx {\lambda\over 2}\left\{
\frac{T^{2}}{12}-\frac{mT}{4\pi }-\frac{ m^{2}}{16\pi ^{2}}\left( \ln
\frac{m^{2}}{(4\pi T)^{2}}+2\gamma -1\right)\right\},
\eeq
where the expansion in the r.h.s. holds when $m\ll T$.
In particular, when $m=0$, we recover the result
$\Pi_2 = \hat m_D^2$, cf. eqs.~(\ref{mpert})--(\ref{mhat}).
This quantity is  generally
dubbed a  ``hard thermal loop'' (HTL), because when $m\ll T$
the dominant
momenta in the one--loop integral (\ref{Sigma2}) are hard ($k\sim T$).

\begin{figure}[t]
\epsfxsize=2cm
\centerline{\epsffile{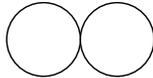}} 
\vspace{3mm}
\caption[a]{The lowest order correction to the thermodynamic potential in
$\phi^4$ theory.}
\label{fig:eye}
\end{figure}

The contribution $\Omega_2$ is
   given by the 2--loop diagram in Fig.~\ref{fig:eye}, and reads:
\beq\label{Omega_2}
\frac{\Omega_2}{V}=\frac{\lambda}{4!}\langle \phi^4\rangle_0\,+\,
\frac{1}{2} \,\delta m^2 \langle \phi^2\rangle_0
= \frac{\lambda}{8}\left(I_T^2(m)-I_0^2(m)\right)\,,
\eeq
where we have also included the contribution of
the mass counterterm and  used
eq.~(\ref{counterterm}) to verify that the  temperature--dependent
ultraviolet divergences cancel out, as they should.
The second term in the last
line in eq.~(\ref{Omega_2}) is divergent, but contributes only to the
vacuum energy density, so that it can be discarded.
For $m=0$, eq.~(\ref{Omega_2}) yields:
\beq\label{P2}
P_2\,=\,-\frac{\lambda}{1152}\,T^4\,=\,
-{15\over 8}\left(\frac{g}{\pi}\right)^2 P_0\,,\eeq
in agreement with eq.~(\ref{Pphi}).

At the next order in the loop expansion, the self--energy receives
contributions from the  2--loop diagrams. 
However, some of these contributions are
{\it infrared divergent} in the massless limit $m=0$.
Consider, e.g., the second diagram displayed in Fig.~\ref{ringPI}. The upper
loop (together with the corresponding mass counterterm, not
shown explicitly in Fig.~\ref{ringPI})
is recognized as the ``hard thermal loop''
computed previously. Thus, this particular 2--loop contribution reads:
\be\label{Pa2l}
\Pi_{2L}^{(a)}\,=\,-\frac{\lambda}{2}\,T\sum_{n}
  \int\frac{{\rm d}^3k}{(2\pi)^3}\,\frac{\hat m_D^2}{(\omega_n^2 + {k}^2)^2}\,.
\ee
The three-dimensional integral over $k$ in eq.~(\ref{Pa2l})
has a linear infrared divergence,
coming from  the term with $\omega_n=0$.
(For the other terms, $\omega_n\ne 0$, the Matsubara frequency
$\omega_n$ acts effectively as an infrared cutoff.)
A non--zero mass $m$ would
cut off this divergence, and give a contribution
$\sim \lambda \hat m_D^2 (T/m) \sim g^4 T^3/m$.
As we shall see, when $m\ll gT$, the real cut-off is not actually the 
mass $m$ itself,
  but the ``thermal mass'' $\hat m_D = gT$ induced by the thermal fluctuations.
Then  $\Pi_{2L}^{(a)}\sim g^4 (T^3/\hat m_D) \sim g^3 T^2$.
This illustrates how nonanalytic terms arise in  eqs.~(\ref{Pphi}) and 
(\ref{mpert}): potential
infrared divergences are cut-off by `thermal masses'  which, through 
their dependence on $g$, modify
the original expansion in powers of the coupling constant. We 
discuss this in more detail in the
next subsection.

\subsection{Order $g^3$ from various resummation schemes}

\subsubsection{The next--to--leading--order thermal mass}
\label{sec:221}

\begin{figure}[htb]
\epsfysize=1.4cm
\centerline{\epsffile{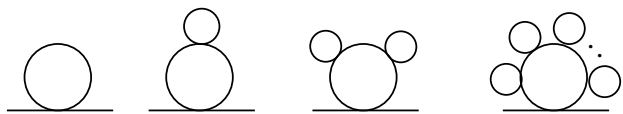}} 
\vspace{3mm}
\caption[a]{Diagrams that are resummed when computing the tadpole in
Fig. \ref{fig:tadpole} with the dressed propagator in eq.~(\ref{hatD}).}
\label{ringPI}
\end{figure}

  The effect
of  a thermal mass is most directly seen by carrying out
perturbative calculations with the dressed propagator
\be\label{hatD}
\hat D(i\omega_n,{\bf k})\,\equiv \,
\frac{1}{\omega_n^2 + {\bf k}^2 + \hat m_D^2}\,.\ee
For the calculation of the tadpole, this corresponds to resumming
the diagrams of bare perturbation theory shown in Fig. \ref{ringPI} 
(each of the inserted tadpoles
in Fig. \ref{ringPI}
is accompanied by a mass counterterm which removes its ultraviolet
divergence in the vacuum, cf. eq.~(\ref{counterterm})).

Since $\hat m_D=gT$, this resummation involves arbitrarily high powers of
$g$. However, the final result can be read from eq.~(\ref{Sigma2})
in which we set $m=\hat m_D=gT$:
\beq\label{g3PI}
\Pi_{\rm tadpole}
\,=\,g^2T^2 \,-\,\frac{3}{\pi}\, g^3T^2 \,+\,{\cal O}(g^4\ln g)\,.\eeq
This formula exhibits the
next--to--leading order contribution to the thermal
mass,  of order $g^3$, consistent with eq.~(\ref{mpert}).
This order--$g^3$ contribution
comes from the term linear in $m$ in the r.h.s. of eq.~(\ref{Sigma2}). It
  thus is related to the non--analyticity of the thermal integral $I_T(m)$
as a function of $m^2$ and to the infrared divergences which occur 
when one attempts to
evaluate this integral after expanding the integrand in powers of $m^2$.
These infrared divergences are easily identified in the sum of the 
diagrams in Fig.
\ref{ringPI} with at least two loops:
\beq\label{Piring}
\Pi_{\rm ring}&=&{\lambda\over 2}\,T\sum_{n}
   \int\frac{{\rm d}^3k}{(2\pi)^3}\,\left\{
\frac{1}{\omega_n^2 + {k}^2 + \hat m_D^2} \,-\,\frac{1}{\omega_n^2 +
{k}^2}\right\}\nn
&=&12g^2\,T\sum_{\omega_n}
   \int\frac{{\rm d}^3k}{(2\pi)^3}\,\frac{1}{\omega_n^2 + {k}^2}\,
\sum_{l\ge 1}\left(\frac{-\hat m_D^2}
{\omega_n^2 + {k}^2}\right)^l\,,\eeq
where the subtracted term is the massless tadpole.
The term with $l=1$ corresponds to the 2--loop
diagram that we have considered separately in eq.~(\ref{Pa2l}).
Each of the terms with $\omega_n^2 > 0$ in eq.~(\ref{Piring})
starts to contribute
at order $g^2 \hat m_D^2\sim g^4$. But the terms with
$\omega_n=0$ are individually infrared divergent for any $l$, and the
divergences cancel only in the sum of all the terms.
In fact, the contribution of order $g^3$
arises from the {\it static} modes alone, i.e.,
the modes with zero Matsubara frequency. If we separate their
contribution in eq.~(\ref{Piring}), we obtain indeed
\beq\label{PI3}
\Pi_{\rm ring}^{(3)}\,=\,12g^2\,T
\int\frac{{\rm d}^3k}{(2\pi)^3}\,\left\{
\frac{1}{{k}^2 + \hat m_D^2} \,-\,\frac{1}{{k}^2}\right\}
\,=\,-\,\frac{3}{\pi}\, g^3T^2,\eeq
where we recognize the order $g^3$ contribution in
    eq.~(\ref{g3PI}).

It is instructive to look at this contribution from another perspective.
As we have argued before, the dominant contribution to $I_T(m)$ in 
eq.~(\ref{Sigma2}) comes from
momenta $k\sim T$. The subleading contribution is
  coming from soft momenta $k\ll T$. To isolate it,
we subtract the hard modes contribution, which amounts to subtract 
$I_T(0)$ from $I_T(m)$ in
eq.~(\ref{Sigma2}), and write:
\beq\label{tad-soft}
  I_T(m)-I_T(0)\,\simeq \, T\int\!\!{d^3k\0(2\pi)^3}\left(\frac{1}
{k^2+ \hat m_D^2}-\frac{1}{k^2}\right)\,=\,
-\frac{\hat m_D T}{4\pi}\,.
\eeq
After multiplication by $\lambda/2=12g^2$, one recognizes the $g^3$ term of
eq.~(\ref{g3PI}). In writing the approximate equality above,
we have used the fact that the dominant contribution in the integral
(\ref{tad-soft}) comes from momenta $k\ll T$, so that we can use
\be\label{BEsoft}
n(\varepsilon_k)\,\approx\,\frac{T}{\varepsilon_k}\,\qquad
{\rm for}\qquad \varepsilon_k\ll T\,.\ee

Note that the above resummation of the thermal mass, focusing on the 
static mode,  hides
an ultraviolet problem that we have ignored so far.
Namely, in order to obtain the finite
result in eq.~(\ref{g3PI}), we had to cancel an ultraviolet divergence,
cf. eq.~(\ref{counterterm}), which now
involves the thermal mass, and is therefore temperature dependent.
The $T$--dependent divergent piece is of order $g^4$ however (since 
proportional
to $g^2\hat m_D^2$),
and thus does not formally affect the order--$g^3$ calculation. But 
it signals that
our  calculation is not complete. The divergence is identified in 
eq.(\ref{Piring}) in
the term  with $l=1$,
  i.e., in the
2--loop diagram in Fig. \ref{ringPI}, after performing the Matsubara sum.
The overall divergence of the
2--loop diagram arises as the product of a vacuum divergence
in the 4--point function subgraph (the lower loop) times the
finite--temperature piece of the upper loop.
Thus, the divergence is removed by the vacuum renormalization
of the 4--point function, i.e., by adding a 1-loop counterterm
to the vertex in the tadpole.

\subsubsection{The plasmon effect in the pressure}

\begin{figure}[htb]
\epsfysize=1.5cm
\centerline{\epsffile{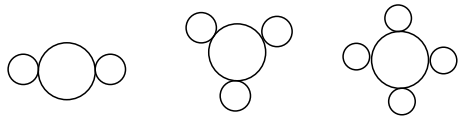}}
\vspace{3mm}
\caption[a]{The ``ring'' diagrams responsible for the order--$g^3$
effect in the pressure.}
\label{ring}
\end{figure}

The contribution of order $g^3$ to the pressure, also
known as the ``plasmon effect'', comes again from resumming diagrams which
are infrared divergent in the na\"{\i}ve diagrammatic expansion.
These are the ``ring'' diagrams in Fig.~\ref{ring}, in which
the central loop is static ($\omega_n=0$),
while each outer loop represents an insertion of $\hat m_D^2$.
This series can be easily summed up, with an infrared finite
result\cite{Kap:FTFT}:
\beq\label{P3}
P_3&=&- {T\0 2}\int\frac{{\rm d}^3k}{(2\pi)^3}\,
\sum_{l\ge 2} \frac{(-1)^{l-1}}{l}\left(\frac{\hat m_D^2}{{k}^2}\right)^l
\nn &=& - {T\0 2} \int\frac{{\rm d}^3k}{(2\pi)^3}\,
\left[\log\left(
1+\frac{\5 m_D^2}{k^2}\right)-\frac{\5 m_D^2}{k^2}\right]=
\frac{\hat m_D^3 T}{12\pi} \,, \eeq
which is the order--$g^3$ contribution in eq.~(\ref{Pphi}).

\subsubsection{Static resummation}

  From the previous examples, we
learned that (a) the occurrence of infrared divergences
in the diagrammatic expansion calls for  resummations and  (b) it 
appears sufficient, in order to
cure these divergences, to resum $\hat m_D^2$ in the static Matsubara 
modes alone.

Since $\hat m_D^2$ summarizes interaction effects, care should be taken
to avoid {\it overcounting} in carrying out high order calculations 
within that scheme.
One way to proceed systematically  is to add and subtract
a mass term for the static modes in the (imaginary--time) Lagrangian:
\beq\label{LE}
{\cal L}&\equiv& {1\over2}(\partial_\tau\phi)^2+{1\over2}(\nbfgrad\phi)^2+
{\lambda\over 4!}\phi^4\nn
&=& {1\over2}(\partial_\tau\phi)^2+{1\over2}(\nbfgrad\phi)^2+
\,{1\over2}\,{\hat m^2_D}\phi^2_0\,+\,{\lambda\over 4!}\phi^4
\,+\Delta {\cal L},\eeq
where $\Delta {\cal L}\equiv -\2 \hat m^2_D\phi^2_0$ and 
$\phi_0({\bf x})\equiv T\int_0^\beta
d\tau \phi({\tau}, {\bf x})$.
This allows for a reorganization of the perturbation theory in which
the tree--level amplitudes are generated by the first three terms in 
eq.~(\ref{LE})
while $\Delta {\cal L}$ is treated
formally as an additional interaction, of order $g^2$, to ensure that the
HTL is not double counted in the static sector.

This procedure preserves a strict correspondence\footnote{This is true,
strictly speaking, only if one uses dimensional regularization
to control the ultraviolet divergences.}
between diagrams (or well identified pieces of it) and powers of $g$.
Thus, it is well suited for weak coupling expansions of static 
quantities, and it is  indeed the
scheme in which the higher--order calculations for the pressure and 
the screening mass alluded to
before have been originally performed (for both scalar
theory\cite{Arnold:1994ps,Parwani:1995zz,Braaten:1995cm} and
QCD\cite{Arnold:1995eb,Zhai:1995ac,Braaten:1996jr}).

\subsubsection{Dimensional reduction}
\label{sec:scdr}

The special role of the static mode $\phi_0$ may be handled in a more 
systematic fashion by
constructing an {\it effective theory} in which the  non--static 
modes $\phi_{\nu\ne 0}$
are `integrated out' in perturbation theory.
Specifically, one can  write
the thermal partition function as the following path--integral:
\beq\label{ZCE0}
Z={\cal N}
\int{\cal D}(\phi_0)\exp\left\{-S_{\rm eff}[\phi_0]\right\}\,,\eeq
where
\beq\label{ZCE1}
\exp\left\{-S_{\rm eff}[\phi_0]\right\}=\,{\cal N}'
\int{\cal D}(\phi_{\nu\ne 0})\exp\left\{
- \int_0^\beta {\rm d}\tau \int{\rm d}^3x \,{\cal L}_E(x)\right\}\,,\eeq
and $S_{\rm eff}[\phi_0]$ is the effective action. Aside from the 
direct classical field
contribution (i.e., the restriction of eq.~(\ref{Lagran}) to the 
static mode $\phi_0$),
this effective action  receives also contributions which,
diagrammatically, correspond to connected
diagrams whose external lines are associated to
$\phi_0$, and  the internal lines are the
propagators of the non-static
modes $\phi_{\nu\ne 0}$. Thus, a priori,  $S_{\rm eff}[\phi_0]$ contains
operators of arbitrarily high order in $\phi_0$, which are
also non-local.
In practice, however, one wishes to expand  $S_{\rm eff}[\phi_0]$ in terms  of
{\it local} operators, i.e., operators with the schematic structure
$a_{m,\,n}\nabla^m \phi_0^n$ with coefficients $a_{m,\,n}$
to be computed in perturbation theory.

It is then useful to introduce an intermediate scale
$\Lambda$ ($\Lambda\ll  T$) which  separates {\it hard}
($k\gtrsim \Lambda$)
and {\it soft} ($k\lesssim \Lambda$) momenta. All the non-static modes, as well
as the static ones with $k \gtrsim \Lambda$ are {\it hard}
(since $K^2\equiv \omega_n^2 +k^2 \gtrsim \Lambda^2$
for these modes), while the static ($\omega_n=0$) modes
with $k\lesssim \Lambda$ are {\it soft}.
Thus, strictly speaking, in the construction of the effective
theory along the lines indicated above, one has to integrate out also
the static
modes with $k \gtrsim\Lambda$. The benefits of this separation of
scales are that ({\it a}) the resulting effective action for the
soft fields can be made
     {\it local} (since the initially non-local amplitudes can be
expanded out in powers of $p/K$, where $p \ll \Lambda$
is a typical external momentum, and $K\gtrsim
\Lambda$ is a hard momentum on an
internal line), and  ({\it b}) the effective theory is now
used exclusively at soft momenta.
This strategy, which consists of integrating out the non-static modes
     in perturbation theory in order to obtain an effective
three-dimensional  theory for the soft static modes, is generally
referred  to as ``dimensional reduction''%
\cite{Ginsparg:1980ef,Appelquist:1981vg,Nadkarni:1983kb,Nadkarni:1988fh,%
Landsman:1989be,Braaten:1995na,Kajantie:1996dw}.

As an illustration
let us consider a massless scalar theory with quartic interactions;
that is, eq.~(\ref{Lagran}) with $m=0$. The  effective action for the 
soft fields takes the form:
\beq\label{EFFLagran}
\!\!S[\phi_0] =\beta {\cal F}(\Lambda)+\int\!{\rm d}^3x \left\{
{1\over 2}(\nbfgrad\phi_0)^2+{1\over 2}M^2(\Lambda)\phi^2_0+
{g^2_3(\Lambda)}\phi^4_0+\cdots\right\}\!,
\eeq
where ${\cal F}(\Lambda)$ is the contribution of the hard modes
to the free-energy, and the dots stand for higher--order, local operators.
We have changed the normalization of the field ($\phi_0\rightarrow
\sqrt{T}\phi_0$) so as to
absorb the factor $\beta$ in front of the effective action. The
parameters of the  effective Lagrangian in eq.~(\ref{EFFLagran}), like
$M^2(\Lambda)$, $g^2_3(\Lambda)$, etc., are computed
in perturbation theory (they involve diagrams in which all the internal
lines are hard), and depend on the
separation scale $\Lambda$, the original
coupling $g$, and  the temperature $T$ (note that the effective 
theory depends on the temperature
only through these parameters). To lowest order in $g$, $g^2_3\approx g^2T$,
and $M\sim gT$, as we shall see shortly.

Note that  the scale
$\Lambda$ acts as an explicit ultraviolet (UV) cutoff for
the loop integrals in the effective theory. Since it is
arbitrary, the dependence on $\Lambda$ of soft loop contributions
must cancel against the dependence on $\Lambda$ of the parameters
in the effective action. Let us verify this cancellation explicitly
in the case of the thermal mass $M$ of the scalar field,
and to lowest order in perturbation theory. To this order, the
mass parameter $M^2(\Lambda)$ in the effective action is obtained by
integrating
over hard momenta within the one-loop diagram in
     fig.~\ref{fig:tadpole} (cf. eq.~(\ref{Sigma})). This gives
\beq\label{MLAM}
M^2(\Lambda)&=&{12g^2}\,T\sum_n
\int\frac{{\rm d}^3k}{(2\pi)^3}\, \frac{(1-\delta_{n 0})+
\theta(k-\Lambda)\delta_{n 0}}{\omega_n^2+k^2}\nonumber\\
&=&{12 g^2}\int\frac{{\rm d}^3k}{(2\pi)^3}\,
\left\{\frac{n(k)}{k}\,+\,\frac{1}{2k}\,-\,\theta(\Lambda-k)
\frac{T}{k^2}\right\}.\eeq
The first two terms within the last integral are the same as
in eq.~(\ref{Sigma}) with $m=0$. The first of them,
involving the thermal distribution, gives
the HTL contribution $\hat m^2_D=g^2T^2$, while the second one,
involving $1/2k$, is removed by the vacuum renormalization.
The third term, involving the $\theta$-function, is easily
evaluated. One finally gets:
\beq\label{MLAM1}
M^2(\Lambda)\,=\,\hat m^2_D\,-\,\frac{6g^2}{\pi^2}\,\Lambda T\,
\equiv\,
{g^2T^2}\left(1-\frac{6}{\pi^2}\,\frac
{\Lambda}{T}\right).\eeq
The $\Lambda$-dependent term above is subleading, by a
factor $\Lambda/T\ll 1$.

The one-loop correction to the thermal mass within the effective
theory is given by the same tadpole diagram, fig.~\ref{fig:tadpole}, 
evaluated with the massive
propagator $1/(k^2+M^2(\Lambda))$,
  coupling constant $g^2_3\approx g^2 T$, and ultraviolet cut-off 
$\Lambda$. We obtain
\beq\label{MDCL1}
\delta M^2(\Lambda) &=&
12g^2T\int\frac{{\rm d}^3k}{(2\pi)^3}\,\Theta(\Lambda-k)\,
\frac{1}{k^2+M^2(\Lambda)}\nonumber\\
&=&\frac{6 g^2}{\pi^2}T\Lambda\left(
1-{\pi M\over 2\Lambda}\,\arctan\,{M\over \Lambda}\right)
\simeq
\frac{6 g^2}{\pi^2}\,T\Lambda
-\frac{3g^2}{\pi} T \hat m_D\,,\eeq
where the terms neglected in the last step
are of higher order in $\hat m_D/\Lambda$ or $\Lambda/T$.

As anticipated, the $\Lambda$-dependent terms cancel in the
sum $M^2\equiv M^2(\Lambda)+\delta M^2(\Lambda)$, which then provides
the physical thermal mass within the present accuracy:
     \beq \label{MTOTS}
M^2\,=\,M^2(\Lambda)+\delta M^2(\Lambda)
\,=\,{g^2T^2}\,-\,\frac{3}{\pi} g^3 T\,,\eeq
which agrees again with eq.~(\ref{g3PI}).

\subsubsection{Screened perturbation theory}

Still another  strategy has been used for taking into account the 
effects of the screening mass to
all orders, rather than perturbatively. It consists of resumming the HTL 
$\hat m_D^2$  in {\it all} the  Matsubara modes, static and non--static (as 
we did in subsection \ref{sec:221}).
It can be formulated via a simple generalization of eq.~(\ref{LE}), viz.
\beq\label{effLagran}
{\cal L}&=& {1\over2}\partial_\mu\phi\partial^\mu\phi\,-\,{1\over2}\,
{\hat m^2_D}\phi^2 - {\lambda\over 4!}\phi^4\,+\,
\Delta {\cal L},\qquad
\Delta {\cal L}\,\equiv\,{1\over2}\,{\hat m^2_D}\phi^2\,,
\eeq
with $\Delta {\cal L}$ to be treated
as a quantity of order $g^2$ in perturbative calculations. This 
scheme is commonly referred to as
{\it screened perturbation theory} (SPT)%
\cite{Karsch:1997gj,Andersen:2000yj}. To appreciate its
virtues, and its limitations, it is again best to work out some examples.

\begin{figure}[tb]
\centerline{\includegraphics[bb=125 250 470 470,width=6.5cm]{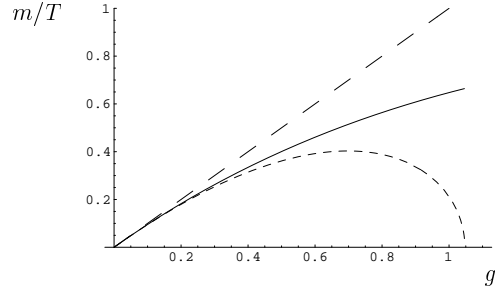}}
\caption{
The one-loop thermal mass in SPT,
$m_D =\Pi_{\rm tadpole}^{1/2}$,   eq.~(\ref{tildePif}),
normalized to the temperature as a function of $g$
(full line). For comparison, we also present the perturbative
estimates to order $g^2$ (long-dashed line)
and $g^3$ (short-dashed line).}
\label{mDPif}
\end{figure}

We have obtained earlier the expression of the tadpole calculated 
with the dressed propagator
(\ref{hatD})
  (cf. eqs.~(\ref{g3PI}) and (\ref{Sigma2})):
\beq\label{tildePif}
\Pi_{\rm tadpole}\,\simeq\,12g^2
I_T(\hat m_D).\eeq
We have seen  that $\Pi_{\rm
tadpole}(\hat m_D=gT)$, when expanded in powers of $g$, reproduce the
correct perturbative result to order $g^3$ (see eq.~(\ref{g3PI})). 
But, as $g$ grows,  it
deviates rapidly from this perturbative result, as can be seen in 
Fig. ~\ref{mDPif}. The poor
convergence of the weak coupling expansion exhibited here is related to that
of the
high--temperature expansion in eq.~(\ref{Sigma2}) which, as we have 
seen, involves in fact non
analytical terms. By including the screening mass in the tree-level 
Lagrangian, and not treating
it a perturbative quantity of order $g$, SPT provides a smooth 
extrapolation to large
values of
$g$, which represents a definite improvement over perturbation theory.

However, in contrast to the static resummation scheme based on eq.~(\ref{LE}),
the calculations within SPT generally involve
temperature--dependent ultraviolet divergences. Of course, these are
compensated by the ``counterterm'' $\Delta {\cal L}$, but  only
in {\it all--order} calculations. At any finite--order,
the compensation holds only to the {\it perturbative}
accuracy\footnote{By the ``perturbative accuracy'' of an all-order result
we understand the highest order in $g$ that is correctly reproduced
by the weak coupling expansion of that result.} of the calculation.
For instance, in order to obtain (\ref{tildePif}),
one has to introduce the mass counterterm $\delta 
m^2=-(\lambda/2)I_0(\hat m_D)$,
which
depends on $T$, via $\hat m_D$. Specifically
(with dimensional regularization, for definiteness, and
$\bar\mu^2=4\pi{\rm e}^{-\gamma}\mu^2$):
\beq\label{counterterm2}
\delta m^2=-12g^2\mu^\epsilon
\int\frac{{\rm d}^{3-\epsilon}k}{(2\pi)^{3-\epsilon}}
{1\over 2\varepsilon_k(\hat m_D)} \simeq \frac{3g^2\hat m^2_D}{4\pi^2}
\left(\frac{2}{\epsilon}+\ln\frac{\bar\mu^2}{\hat m^2_D}+1\right),
\eeq
where $\varepsilon_k(\hat m_D)^2=k^2+\hat m_D^2$. This formula makes 
explicit the fact that the
$T$--dependent divergence  is of order ${\cal O}(g^4)$,
and thus lies beyond the perturbative accuracy
of this one--loop calculation, which is ${\cal O}(g^3)$.

As further illustration of the difficulties with SPT, let us compute 
the pressure
to 2--loop order, and compare the result
with its perturbative expansion to order $g^3$.

To one--loop order in the effective theory, the
thermodynamic potential is given by
eq.~(\ref{O0}) with $m\to \hat m_D$. That is:
\beq\label{O0eff}
\frac{ \Omega_0}{V}&=&
\2\,T\sum_{n}
   \int\frac{{\rm d}^3k}{(2\pi)^3}\,\ln (\omega_n^2 + {\bf k}^2 + \hat m_D^2)
\nn
&=&\2
\int\! {d^3k\0(2\pi)^3}\,\varepsilon_k(\hat m_D)\,\,-\,\,P_0(\hat m_D)
\eeq
with $P_0(m)$ given by  eq.~(\ref{P0mexp}):
\beq\label{Pm_D}
P_0(\hat m_D)\,=\, \frac{\pi^2T^4}{90}\,\left\{
1-{30\over 8}\left(\frac{g}{\pi}\right)^2
+{15\over 2}\left(\frac{g}{\pi}\right)^3 \,+\,{\cal O}(g^4\ln g)\right\}.
\eeq
This overincludes the order--$g^2$ effect by a factor
of 2 (compare to eq.~(\ref{Pphi})),
but includes correctly the term of order $g^3$ (the fact
that the plasmon effect is included correctly is somewhat accidental:
If we were to resum not just the leading--order thermal
mass $\hat m_D^2$, but also the NLO correction to it,
$\Delta  m_D^2 = -(3/\pi)g^3T^2$, one would modify the
order--$g^3$ content of $ \Omega_0$).
The first piece in eq.~(\ref{O0eff}), i.e., the
sum of the ``zero-point energies'', is also UV divergent, and the 
divergent terms depends on $T$
since they depend on $\hat m_D$.
To eliminate these divergences, we may
introduce a temperature--dependent counterterm:
\be\label{ct2}
\frac{\delta  \Omega_0}{V}\,=\,-\2
\int\! {d^3k\0(2\pi)^3}\,\varepsilon_k(\hat m_D)\,=\,-\2
\int\! {d^4k\0(2\pi)^4}\,\ln({k_0^2 + {\bf k}^2 +\hat m_D^2})\,,\ee
whose dependence upon $T$ starts already at order $g^2$.
(This dependence can be pushed to order $g^4$, though, by using dimensional
regularization.)
This lowest order calculation reveals another unsatisfactory feature 
of SPT: the tree--level
pressure calculated with the Lagrangian (\ref{effLagran}) fails to 
reproduce  the lowest--order
effect of the interactions.

Of course these drawbacks are partially corrected when we move to the 
next order, that is  to
two--loop order, which we do now. At this level, there are three 
contributions: the two--loop
diagram in Fig.~\ref{fig:eye}, and two one--loop diagrams involving
mass counterterms, namely $\Delta {\cal L}$
in the Lagrangian (\ref{effLagran}), whose role is to correct for
overcounting, and the ultraviolet counterterm $\delta m^2$, that has been
introduced before, in eq.~(\ref{counterterm2}).
Altogether:
\be\label{O2l}
\frac{  \Omega_2}{V}\,=\,
\frac{\lambda}{8}
\left[I(\hat m_D)\right]^2 \,+\,\frac{1}{2} \Big(\delta m^2
- \hat m_D^2\Big)I(\hat m_D)\,.\ee
It is convenient to combine the last counterterm with the tree--level
expression $ \Omega_0$ (written as in the first line of
   eq.~(\ref{O0eff})):
\bea\label{O22}
&&\frac{ \Omega_0}{V}\,-\,\frac{\hat m_D^2}{2}\,I(\hat m_D)
\nonumber\\&&=\,
\2\,T\sum_{n}
   \int\frac{{\rm d}^3k}{(2\pi)^3}
\Bigl\{\ln (\omega_n^2 + {\bf k}^2 + \hat m_D^2)
-\frac{\hat m_D^2}{\omega_n^2 + {\bf k}^2 + \hat m_D^2}\Bigr\}\,.\eea
Clearly, the main effect of $\Delta {\cal L}$ is to cancel out the
whole contribution of order $g^2$ that was originally present
in the tree--level pressure. This includes the ``anomalous''
order--$g^2$ effect within $P_0(\hat m_D)$, and the divergent
contribution of order $g^2$ that was eliminated by the
counterterm  (\ref{ct2}) at one-loop order. The  terms  in 
eq.~(\ref{O2l}) can be simplified as
follows:
\beq\label{O230}
\frac{\lambda}{8}
\left[I(\hat m_D)\right]^2 \,+\,\frac{\delta m^2}{2}\,I(\hat 
m_D)=\frac{\lambda}{8}\left[
I_T(\hat m_D)\right]^2\,-\, \frac{\lambda}{8}\left[
I_0(\hat m_D)\right]^2\,,
\eeq
(cf. eq.~(\ref{Omega_2})).
The last term is divergent and depends
on the temperature, via  $\hat m_D$, but this
dependence counts only to order $g^4$, or higher.
The order--$g^2$ is obtained by replacing $I_T(\hat m_D)$ by $I_T(0)$ 
in the first term in
eq.~(\ref{O230}).
The order  $g^3$ receives contributions both from  eq.~(\ref{O22}) 
(as the lowest order
correction beyond the ideal gas result), and from  eq.~(\ref{O230})
(as the next--to--leading order contribution). In both cases,
  the integration involves  soft momenta only. Thus,
we can restrict ourselves to the static mode $\omega_n=0$ in
eq.~(\ref{O22}), and get:
\be\label{O24}\2\,T
   \int\frac{{\rm d}^3k}{(2\pi)^3}
\left\{\ln \left(1 +\frac{\hat m_D^2}{k^2}\right)
\,-\,\frac{\hat m_D^2}{k^2 + \hat m_D^2}\right\}\,,\ee
while in  eq.~(\ref{O230}) we may proceed as follows:
\beq \label{O25}
\frac{\lambda}{8}\left(
I_T(\hat m_D)\right)^2\,-\, \frac{\lambda}{8}\left(
I_T(0)\right)^2
\simeq \frac{\lambda}{4}
I_T(0) \,
\left(I_T(\hat m_D)-I_T(0)\right)\nn
   \,\simeq\,\hat m_D^2\,\2\, T\int\! {d^3k\0(2\pi)^3}\left(\frac{1}
{k^2+ \hat m_D^2}-\frac{1}{k^2}\right)
\eeq
where we have also used eq.~(\ref{tad-soft}).
Clearly, after adding the two contributions in eqs.~(\ref{O24})
and (\ref{O25}), we are left with the standard expression for $P_3$,
eq.~(\ref{P3}).

Now, all the remaining ultraviolet divergences in eqs.~(\ref{O22})
and (\ref{O230}) are to be cancelled by appropriate counterterms.
One could be satisfied by the fact that the $T$--dependence of these 
counterterms starts at order
$g^4$, which is beyond the perturbative accuracy of
the present 2--loop calculation (namely, ${\cal O}(g^3)$).
But leaving temperature dependence in UV counterterms is somewhat 
unphysical and would cause
problems in  {\it non--perturbative} calculations. Moreover,
ambiguities arise from the choice of  the subtraction scheme:
One could e.g.~either choose to subtract all the $T$-independent
terms of the thermal pressure before identifying $\hat m_D$
with a thermal mass, or one may only subtract pole terms
minimally on the grounds that also the terms not explicitly
dependent on $T$ become so after making $\hat m_D$ temperature-dependent.

Some of these drawbacks may actually be absorbed by the flexibility in 
the choice of the mass
parameter $\hat m_D$ in (\ref{effLagran}). Obviously, the results of 
SPT would be independent
of this mass if they were obtained in an all order calculation.
But for a finite--order calculation, the choice matters. One may try 
to optimize perturbation
theory by a principle of minimal sensitivity resulting  in a self--consistent
gap equation\cite{Karsch:1997gj,Andersen:2000yj}.
This has been shown to  reduce greatly (though not completely)
the dependence on the subtraction procedure%
\cite{Rebhan:2000uc} (see also below, Fig.~\ref{figscP}).
Indeed, in  Ref.~\cite{Andersen:2000yj} the calculation of the pressure
of scalar $\phi^4$ theory has been extended
to 2--loop order for the screening mass and  3--loop order for the pressure
(meaning a perturbative accuracy of order $g^5$ in both cases),
with results showing an excellent apparent convergence: i.e., the
difference between the 2--loop and 3--loop results for the pressure
remains rather small for all $g\lesssim 1$.

\subsection{Self--consistent resummation}

\subsubsection{Skeleton expansion for thermodynamical potential} 

A systematic way to take into account in the calculation of the 
thermodynamics, screening effects,
or more generally propagator renormalisations, is to
use the representation%
\cite{Luttinger:1960,Baym:1962,DeDominicis:1964,Cornwall:1974vz}
of the thermodynamic potential $\Omega=-PV$
in terms of the full propagator $D$:
\be
\label{LW}
\b \Omega[D]=-\log Z=\2 \Tr \log D^{-1}
-\2 \Tr \Pi D+\Phi[D]\,,
\ee
where $\Tr$ denotes the trace in configuration space,
$\b=1/T$, and $\Phi[D]$ is the sum of the {\it two--particle irreducible} (2PI)
``skeleton'' diagrams with no external legs:
\be\label{skeleton}
-\Phi[D]=
\epsfxsize=6cm
\epsfbox[50 392 550 410]{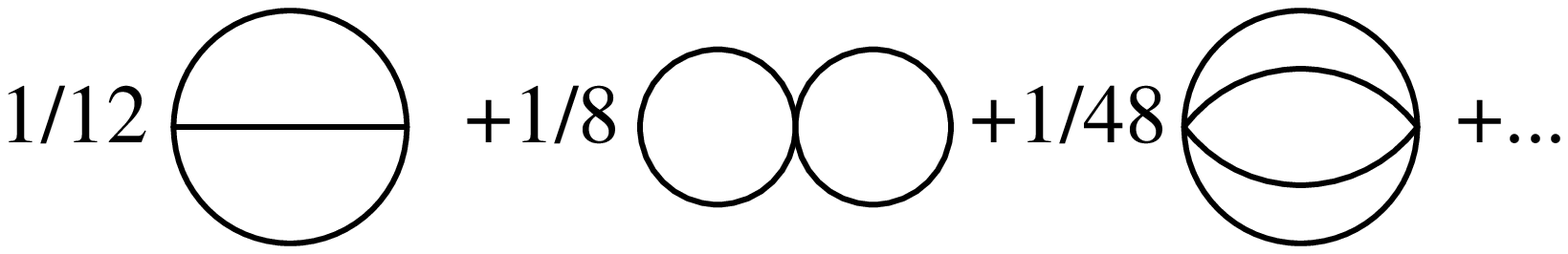}\phantom{\Bigg|_1}
\ee
(where in the $\phi^4$ example the first diagram is absent of course).
The full propagator $D$ is expressed in terms of the proper 
self--energy $\Pi$ by Dyson's equation
($D_0$ is the free propagator):
\beq\label{Dyson}
D\,=\,\frac{1}{D^{-1}_0+\Pi}\,,
\eeq
and
$\Pi$ itself, which is the sum of the one--particle irreducible (1PI) 
diagrams with
two external lines, is obtained from  $\Phi[D]$ by:
\be\label{PhiPi}
\d\Phi[D]/\d D\,=\,\2\Pi\,.
\ee
By using this relation, one can check that $\Omega[D]$ is stationary 
under variations of $D$
(at fixed $D_0$) around the physical propagator:
\be\label{selfcons}
{\d\Omega[D] / \d D}\,=\,0\qquad {\rm for}\qquad D=(D^{-1}_0+\Pi)^{-1}\,.
\ee
We shall usually refer to eq.~(\ref{Dyson}) in which $\Pi$ is given 
in terms of $D$ by
eq.~(\ref{PhiPi}) as a `gap equation'.

An explicit expression for $\Omega$ is obtained by performing
the  summations over the Matsubara frequencies
in eq.~(\ref{LW}), using standard
contour integration techniques. One obtains:
\be\label{Omega(D)}
\Omega/V=\int\!\!{d^4k\0(2\pi)^4}\, n(\omega)\left(\Im \log(-\omega^2+k^2+\Pi)
-\Im\Pi D\right)+T\Phi[D]/V
\ee
where $n(\omega)=1/({\rm e}^{\beta \omega}-1)$, and the imaginary parts
are defined with retarded prescription. For instance:
\beq
\Im D(\omega,k)\equiv \Im D(\omega+i\epsilon,k)=\frac{\rho(\omega,k)}{2}\,,
\eeq
where $\rho(\omega,k)$ is the spectral function, which enters the
following representation of the analytic propagator $D(\omega,k)$,
valid for frequencies $\omega$ off the real axis:
\be
D(\omega,k)=\int_{-\infty}^\infty {dk_0\02\pi}{\rho(k_0,k)\0k_0-\omega}\,.
\ee


The stationarity condition (\ref{selfcons})
can be used as a variational principle to deduce non--perturbative
approximations for the  physical propagator.
A convenient approximation scheme of this type
is the {\it self--consistent} (or ``$\Phi$--derivable'')
approximation\cite{Baym:1962}
obtained by selecting a class of skeletons in
$\Phi[D]$ and calculating $\Pi$ from eq.~(\ref{PhiPi}). We shall see 
also that the stationarity
property (\ref{selfcons}) brings simplifications in the calculation 
of the entropy.

\subsubsection{A simple model}
\label{secsimplemodel}

As a simple illustration of the general formalism of the previous 
subsection,  we consider here the
2--loop approximation to
$\Phi$, given by the first skeleton in the r.h.s. of
eq.~(\ref{skeleton}).\footnote{The self-consistent solution to 3-loop
order has been worked out using mass expansions in
Ref.~\cite{Braaten:2001en
} with satisfactory
numerical results. However, the negative conclusion therein concerning
renormalizability of $\Phi$-derivable
approximations beyond two-loop order has since been
refuted by
Refs.~\cite{vanHees:2001ik,
Blaizot:2003br}.}
The corresponding skeleton for $\Pi$ is the dressed tadpole (see
Fig. \ref{fig:tadpole}),
and is local. Thus, the self--consistent self--energy is simply a mass,
and we shall write $\Pi\equiv m^2$. (The vacuum mass is set to zero.)
The gap equation (\ref{PhiPi}) reads then:
\beq \label{gapp}
m^2\,=\,{\lambda\over 2} I(m).
\eeq

In contrast to the previous
renormalization of eq.~(\ref{tildePif}) (recall eq.~(\ref{counterterm2})),
it is possible to renormalize  eq.~(\ref{gapp}) without
introducing a thermal counter\-term. This is done through coupling 
constant renormalization. To
proceed, we first  rewrite
eq.~(\ref{gapp}) as (with the same notations as in eq.~(\ref{counterterm2})):
\beq\label{I(m)regul}
\mu^{\epsilon}
m^2\,=\,-\lambda_0\,\frac{m^2}{32\pi^2}
\left(\frac{2}{\epsilon}+\log\frac{\bar\mu^2}{m^2}+1\right)
+\,\lambda_0\int\!\!{d^3k\0(2\pi)^3}\,\frac{n(\varepsilon_k)}
{2\varepsilon_k}\,+{\rm O}(\epsilon),
\eeq
and relate  $\lambda_0$ to
the renormalized coupling $\lambda$ by:
\beq\label{RENL}
\frac{1}{\lambda}=\frac{\mu^{\epsilon}}{\lambda_0}+\frac{1}{16\pi^2\epsilon}.
\eeq
Then, the equation takes the following, manifestly finite, form
(for $\epsilon\to 0$):
\beq\label{GAP2}
m^2\,=\,\frac{\lambda}{2}\int\frac{{\rm d}^3k}{(2\pi)^3}\,
\frac{n(\varepsilon_k)}{\varepsilon_k}\,+\,
\frac{\lambda  m^2}{32 \pi^2}\left(\log \frac{m^2}{\bar\mu^2}
\,-1\right).
\eeq
One should stress a subtle point in the procedure: expressing the bare coupling
  $\lambda_0$ in terms
of the renormalized one $\lambda$ according to eq.~(\ref{RENL})
is not sufficient to render finite the r.h.s. of eq.~(\ref{I(m)regul})
for  {\it arbitrary} values of $m^2$, but only for these values  which
satisfies the finite equation (\ref{GAP2}).

Note that the 1--loop correction to the 4--point function that is considered
here is iterated only in one channel among the three possible ones.
Accordingly, the $\beta$--function for the renormalized coupling constant
(\ref{RENL}):
\be\label{RGSC}
\frac{{\rm d}\lambda}{{\rm d}\log \bar\mu}\,=\,\frac{\lambda^2}{16\pi^2},
\ee
(this follows by noticing that $\lambda_0$ is independent of $\bar\mu$
in eq.~(\ref{RENL})) is only one third of the lowest-order perturbative
$\beta$-function for this scalar field theory (see eq.~(\ref{renorm0})).
This is no actual fault since the running
of the coupling affects the thermodynamic potential only at
order $\lambda^2$ which is beyond the perturbative accuracy of
the 2-loop $\Phi$-derivable approximation. (In order to see the
correct one-loop $\beta$-function, the approximation
for $\Phi$ would have to be pushed to 3-loop order.)
By using eq.~(\ref{RGSC}), one can check that
the solution $m^2$ of eq.~(\ref{GAP2}) is independent of $\mu$.

By using the previous formulae, one can easily compute the
pressure\cite{Drummond:1997cw}:
\beq\label{PPHI}
P=-T\int\frac{{\rm
d}^3k}{(2\pi)^3}\,\log(1-{\rm e}^{-\beta\varepsilon_k})
+{m^2\02}\int\!\!{d^3k\0(2\pi)^3}\,\frac{n(\varepsilon_k)}
{2\varepsilon_k}\,+{m^4\0128\pi^2}\,,
\eeq
which differs from the pressure $P_0(m)$ of an ideal gas of
massive particles, eq.~(\ref{P0mexp}),
by the last two terms in the r.h.s.

\begin{figure}[tb]
\centerline{\includegraphics[bb=70 200 540 540,width=6.5cm]{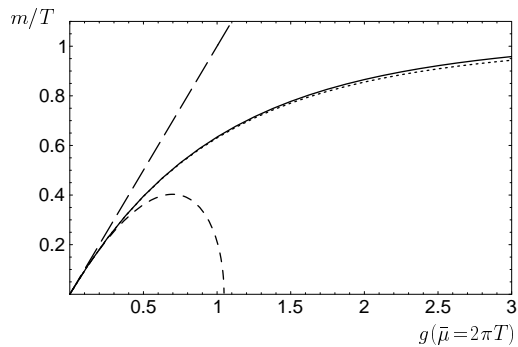}}
\caption{Numerical solution of the mass gap equation (\protect\ref{GAP2})
with $\bar\mu=2\pi T$
in comparison with the perturbative result to order $g^2$ and
$g^3$ for $m$ (long and medium dashed lines, resp.). The short-dashed
line just below the full line corresponds to the quadratic
gap equation (\protect\ref{gapsoft}) following from the
requirement of only soft self-consistency.
}
\label{figscm}
\end{figure}

\begin{figure}[tb]
\centerline{\includegraphics[bb=70 200 540 540,width=6cm]{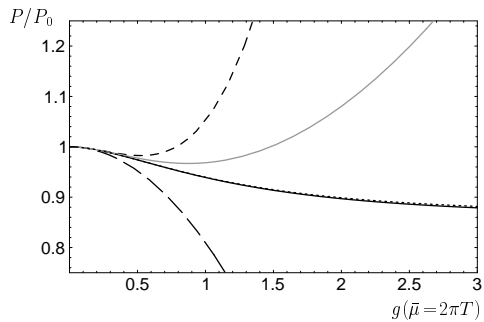}}
\caption{The 2-loop $\Phi$-derivable pressure (\protect\ref{PPHI})
in comparison with the perturbative results to order $g^2$ and
$g^3$ (long and medium dashed lines, resp.). The short-dashed
line just above the full line corresponds to the pressure
evaluated perturbatively for hard modes and self-consistently
only for the soft modes, eqs. (\protect\ref{O2eff}) and
(\protect\ref{LW3Dm}) with (\protect\ref{gapsoft}).
SPT at 2-loop order turns out to coincide with the
2-loop $\Phi$-derivable given by the full line, when the
former is subtracted minimally, but gives different results
(grey full line) when subtracted ``maximally'' (see text).
}
\label{figscP}
\end{figure}

By construction, the 2--loop $\Phi$--derivable approximation is perturbatively
correct to order $g^3$, and, indeed, it can be easily verified that 
the weak coupling
expansions of eqs.~(\ref{GAP2}) and (\ref{PPHI}) include the
correct perturbative effects of order $g^2$ and $g^3$ (together
with incomplete contributions of higher orders). But the complete,
self--consistent results, as obtained by numerical evaluation in
eqs.~(\ref{GAP2}) and (\ref{PPHI}), show a much better behaviour than the
respective perturbative expansions to order $g^3$, in the sense of being
monotonic with $g$ (unlike the perturbative approximants of order $g^3$),
and showing saturation at very large $g\gtrsim 2$ (in contrast to the
order--$g^2$ estimates). This behaviour is illustrated in Figs. \ref{figscm}
and \ref{figscP} for the self-consistent mass (\ref{GAP2})
and pressure (\ref{PPHI}), respectively.

Also shown in Fig.~\ref{figscP} is the result of SPT at two-loop
order when the mass parameter is determined by a principle of
minimal sensitivity. It turns out that for a minimal subtraction
of the additional UV divergences of SPT and identification of
the corresponding regularization mass scale with that used
in the ($\overline{\mbox{MS}}$) renormalization of the coupling
constant, the result of SPT coincides with that of the
2-loop $\Phi$-derivable approximation. However, if the
additional UV divergences of SPT are subtracted differently,
this coincidence no longer occurs. As an example, in Fig.~\ref{figscP}
the grey full line displays the
SPT result with a ``maximal'' subtraction
of the additional UV divergences where $P(T)$ is made finite
by subtracting $P(0)$ when treating $\hat m_D$ as
temperature-independent (since it is just a parameter in the
Lagrangian), and only afterwards turning it temperature-dependent
by fixing $\hat m_D$ by a variational principle.
Although (for $N=1$ scalar theory) one cannot decide how
the exact result would look like, the result appears to be
less satisfactory in that it exceeds the free pressure
for $g>1.5$. The prescription that has been used in the
application of SPT to QCD is in fact that of a minimal
subtraction of the additional UV divergences of SPT%
\cite{Andersen:1999fw,
Andersen:2002ey,Andersen:2003zk}.

We have previously argued that the expected accuracy of perturbation 
theory is a priori different
in the soft and the hard sectors.
Within the present solvable model, it is possible to
verify this argument explicitly. To this aim, we shall reformulate 
the self-consistent
calculation of the thermal mass and the pressure in an approximate 
way, exploiting the distinction
between the hard and soft sectors. That is, we shall build on the 
strategy of effective theories to
obtain perturbatively, through a calculation in the hard sector,  the 
coefficients of an effective
Lagrangian for the soft sector. We shall then look for an ``exact''
solution in the soft sector,
which amounts here to use the two-loop self-consistent approximation. 
As we shall see, this
strategy, which mimics that used in calculations based on dimensional 
reduction,  turns out to be
remarkably accurate. It also provides a clearer physical 
justification for approximations that have
been motivated  previously by other considerations.

Thus, we shall approximate the fully self--consistent 2--loop result
for the pressure, eq.~(\ref{PPHI}), by $P=P_{(2)} +P_{\rm s}$, where
$P_{(2)}$ is the perturbative result to order $g^2$, which is
due to the hard modes alone:
\be
\label{O2eff}
P_{(2)}\,=\,\frac{\pi^2}{90}\, T^4\Bigg[
1-{15\over 8}\left(\frac{g}{\pi}\right)^2\Bigg].\ee
and $P_{\rm s}$ is obtained via a 2--loop  self--consistent calculation
within the effective three--dimensional theory with Lagrangian 
(\ref{EFFLagran}).
To the order of interest, one can set $M^2\approx \hat m_D^2$ and
$g_E^2\approx g^2T$ in eq.~(\ref{EFFLagran}), and neglect all the 
other interaction vertices.

The corresponding free--energy functional reads then:
\bea
\label{LW3D}
\frac{\b \Omega_{\rm s}[\Pi_{\rm s}]}{V}&=&\2 \int_k 
\left\{\ln(k^2+\hat m_D^2+\Pi_{\rm s})
-\frac{\Pi_{\rm s}}{k^2+\hat m_D^2+\Pi_{\rm 
s}}\right\} \nonumber\\
&&+\frac{\lambda}{8}\left(
\int_k \frac{1}{k^2+\hat m_D^2+\Pi_{\rm s}}\right)^2,
\eea
where $\int_k$ denotes the momentum integral in
$3-d$ dimensions. Note that, as compared to Sect.~\ref{sec:scdr}, 
here we find it more
convenient to use dimensional regularization, with the
scale $\mu$ chosen as the separation
scale $\Lambda$ between soft and hard momenta. This has the advantage that
all the power--like divergences 
in integrals like those in eq.~(\ref{LW3D}) are now set to zero,
so, conversely, the parameters of the effective Lagrangian are independent
of $\Lambda$ to the accuracy of interest\cite{Braaten:1995cm}.

In  eq.~(\ref{LW3D}), $\Pi_{\rm s}$ is the scalar self--energy within the $3-d$
effective theory, and is determined from a gap equation which follows 
by demanding
  $\Omega_{\rm s}[\Pi_{\rm s}]$ to be stationary with respect to $\Pi_{\rm s}$:
  \be
\label{Pi3}
\Pi_{\rm s}\,=\,\frac{\lambda}{2} \,T\int_k \frac{1}{k^2+\hat 
m_D^2+\Pi_{\rm s}}\,,\ee
or, equivalently, with $m^2\equiv \hat m_D^2+\Pi_{\rm s}$,
\be\label{gapsoft}
m^2_{\rm s}\,=\,\hat m_D^2 - \frac{3}{\pi} g^2 T m_{\rm s}\,.\ee
This is a second-order algebraic equation for $m_{\rm s}$, whose r.h.s. is
recognized as the expansion of the integral in the general gap equation in
four dimensions, eq.~(\ref{gapp}), up to terms linear in $m/T$.
Note that, to this accuracy, there is no ultraviolet divergence at all,
so the coupling constant $\lambda$ (or $g^2$) in the equations above is
the renormalized one.

The free energy (\ref{LW3D}) can be easily evaluated in terms of the
solution $m_{\rm s}$ to the above gap equation. One obtains ($P_{\rm 
s}=-\Omega_{\rm s}/V$):
\be
\label{LW3Dm}
P_{\rm s}
\,=\,\frac{T m_{\rm s}^3}{12\pi}  - \frac{T m_{\rm s}}{8\pi} (m_{\rm 
s}^2-\hat m_D^2)
-\,\frac{3}{16}\left(\frac{g}{\pi}\right)^2 T^2 m_{\rm s}^2\,.\ee
Note that this contribution starts at order $g^3$, but includes terms
of all higher orders, via $m_{\rm s}$.
The sum $P=P_{(2)}+P_{\rm s}$ of eqs.~(\ref{O2eff}) and (\ref{LW3Dm})
should be compared to the fully self--consistent
result in eq.~(\ref{PPHI}). Clearly, these two approximations
coincide up to order $g^3$, but deviate in higher orders, with the deviations
restricted, however, to contributions due to the {\it hard} modes.
In Figs. \ref{figscm} and \ref{figscP},
we also compare the numerical solutions to
the full gap equation (\ref{GAP2}) and its ``soft''
approximation (\ref{gapsoft}),
and also the corresponding predictions for the pressure.
We see that these
predictions are extremely close to each other, for both $m$ and $P$,
which confirms that the higher order effects associated with
the hard modes are small indeed.

\subsubsection{The entropy}

We have mentioned already that the stationarity property 
(\ref{selfcons}) allows one to simplify the
calculation of the {\it entropy}:
\be
{\cal S}=-{\6(\Omega/V)/\6T}\,
\ee
or, more generally, of any first partial derivative of the pressure.
Indeed, the thermodynamic potential (\ref{Omega(D)}) depends on the
temperature through the statistical factors $n(\omega)$ and the 
propagator, or equivalently
its spectral function $\rho$.
Because of  eq.~(\ref{selfcons}) the
temperature derivative of the spectral
density cancels
out and one obtains%
\cite{Riedel:1968,Vanderheyden:1998ph}:
\bea\label{Ssc}
{\cal S}&=&-\int\!\!{d^4k\0(2\pi)^4}{\6n(\omega)\0\6T} \Im \log D^{-1}(\omega,k) \nn
&&+\int\!\!{d^4k\0(2\pi)^4}{\6n(\omega)\0\6T} \Im\Pi(\omega,k) \Re D(\omega,k)+{\cal S}'
\eea
with
\be\label{SP0}
{\cal S}'\equiv -{\6(T\Phi)\0\6T}\Big|_D+
\int\!\!{d^4k\0(2\pi)^4}{\6n(\omega)\0\6T} \Re\Pi \Im D.
\ee
Loosely speaking, the first two terms in eq.~(\ref{Ssc}), which have a
one--loop  structure,
represent the entropy of `independent quasiparticles', while ${\cal S}'$
accounts for a residual interaction among these quasiparticles%
\cite{Vanderheyden:1998ph} (see also the discussion below).
For 2--loop\footnote{For the scalar field
theory with $\phi^4$ self--interactions, the
property (\ref{Sprime}) holds also for the 3--loop skeleton in
eq.~(\ref{skeleton}).} skeletons we have
the following additional simplification%
\cite{Riedel:1968,Vanderheyden:1998ph,Blaizot:2000fc}:
\beq\label{Sprime}
{\cal S}'=0,
\eeq
a property which holds
independently of the self-consistency condition (\ref{selfcons}).

The formula (\ref{Ssc}), together
with ${\cal S}'=0$,
makes the entropy a privileged quantity to study the thermodynamics of
ultrarelativistic plasmas. Let us discuss some of the attractive features.

First, we note that, for a finite function $\Pi(\omega,k)$,
the integrals in  eq.~(\ref{Ssc}) with ${\cal S}'=0$ are
manifestly ultraviolet finite, since ${\6n/\6T}$
vanishes exponentially for both $\omega\to\pm\infty$.
This reduces the problem of the UV renormalization of the
2--loop $\Phi$--derivable approximation for the entropy
to that of the renormalization
of the gap equation for the 1--loop
self-energy. In general, this is still a difficult task
(because this is a non--perturbative approximation), but as
we have seen in the previous subsection, progress can be made
at least in specific cases. In particular, it has been recently
demonstrated, in Refs.~%
\cite{vanHees:2001ik,
Blaizot:2003br},
that for scalar field theories at least,
the self--consistent approximations are {\it renormalizable} to all orders,
i.e., they can be made finite by adding 
temperature-independent (vacuum)
counterterms to the Lagrangian,
which are themselves determined in a self--consistent way.

Furthermore, as already mentioned, the entropy has a more
direct quasiparticle interpretation than the pressure.
To see this more explicitly, use the following identity:
\beq
\Im \log D^{-1}(\omega,k)=\arctan\left(
\frac{\Im \Pi}{\Re D^{-1}} \right)-\pi\epsilon(\omega)\theta(-\Re D^{-1}),
\eeq
to rewrite ${\cal S}$ as
${\cal S}= {\cal S}_{pole}+{\cal S}_{damp}$,
with
\beq\label{Spole}
{\cal S}_{pole}&=&\int\!\!{d^4k\0(2\pi)^4}{\6n(\omega)\0\6T} 
\pi\epsilon(\omega)
\theta(-{\Re}D^{-1}(\omega,k))\nonumber\\
    &=& \int\!\!{d^3k\0(2\pi)^3}\Bigl\{(1+n_k)\log (1+n_k) \,-\,
n_k\log n_k\Bigr\},
\eeq
where $n_k\equiv n(\varepsilon_k)$, with $\varepsilon_k$ solution
of $\Re D^{-1}(\omega=\varepsilon_k,k)=0$. This is the on-shell energy of the
dressed single--particle excitations in the system, or `quasiparticles'.
The quantity ${\cal S}_{pole}$ is the entropy of a system of
`non-interacting' quasiparticles, while the  quantity
\beq
{\cal S}_{damp} = \int\!\!{d^4k\0(2\pi)^4}{\6n(\omega)\0\6T} \left[{\rm
Im}\Pi\Re D- \arctan\left(
\frac{\Im \Pi}{\Re D^{-1}} \right)\right],
\eeq
which vanishes when $\Im \Pi$ vanishes, is a contribution coming from
the continuum part of the quasiparticle spectral weights. We see that
for the entropy, unlike for the pressure,
the main effects of the interactions
come via the change in the spectrum of the excitations, and therefore
can be taken into account by simply dressing the propagator.

The above formulae, which are generic (i.e.,
they hold for any field theory, including QCD), become even simpler
when the 2--loop self--consistent approximation is applied to
the scalar theory with $\phi^4$ self--interactions. Then, the
self--energy reduces to a local mass term,
as we have seen in the previous subsection, and therefore
${\cal S}={\cal S}_{pole}$, with ${\cal S}_{pole}$ given by eq.~(\ref{Spole})
with $\varepsilon_k^2=k^2 +m^2$ and $m$ determined by eq.~(\ref{GAP2}).
That is, in this approximation, the entropy of the interacting scalar
gas is formally identical to the entropy of an ideal gas of massive
bosons, with mass $m$. As emphasized after eq.~(\ref{PPHI}),
such a simple interpretation does not hold for the pressure.

\subsubsection{Approximately self--consistent calculations of the entropy}

In this subsection, the calculation of the self--consistent 2--loop
entropy will be considered from a new perspective, which will suggest
a different strategy for implementing the self--consistency
in an approximate way. This strategy will be especially useful in view
of subsequent applications to QCD, where a
fully self--consistent calculation seems prohibitively difficult, and not
even desirable (see Sect. \ref{sec:htlresummed}).

It is first instructive to see how the perturbative effects of order $g^2$
and $g^3$ get built within the self--consistent 2--loop entropy, as given by
(cf. eq.~(\ref{Spole})):
\beq \label{Sdem}
{\cal S}(m)&=&-\int\!{d^4k\0(2\pi)^4}{\6n(\omega)\0\6T}\Im\log (k^2-\omega^2+m^2)\nn
&=& \frac{2\pi^2 T^3}{45}-\frac{m^2
T}{12}+\frac{m^3}{12\pi}\,+{\cal O}(m^4/T).
\eeq
The first two terms in the expansion in the second line come from integrating
over hard momenta ($k\sim T$) within the loop integral in the first line,
while the third term, non--analytic in $m^2$, comes from the soft loop
($k\sim m$).

The expansion in eq.~(\ref{Sdem}) must be considered simultaneously with the
perturbative expansion of the solution $m$ to the gap equation (\ref{GAP2}).
To order $g^3$, this is the same as the standard perturbative expansion,
eq.~(\ref{g3PI}): $m^2\simeq g^2T^2(1- 3g/\pi)$.

If one replaces $m$ in eq.~(\ref{Sdem})
by the LO thermal mass $\hat m_D=gT$,
then the resulting approximation for ${\cal S}$
reproduces the perturbative effect of order $g^2$,
but underestimates the correction of order $g^3$ by a factor of 4.
But to order $g^3$, the thermal mass itself must be corrected, as
discussed above. By replacing $m^2\to \hat m_D^2 + \delta m_D^2$
in the second order term of eq.~(\ref{Sdem}),
the correction $\delta m_D^2$ to the thermal mass generates
the 3/4 of the plasmon effect missing.
To conclude, 
the order--$g^3$ contribution
to the entropy comes up in a subtle way: 1/4 of it comes directly
from the {\it soft} loop in eq.~(\ref{Sdem}), with $m$ replaced by  the
LO {\it screening} mass $\hat m_D$, while the other 3/4
comes from the {\it hard} loop, via the NLO correction to the {\it
dispersion relation} of the hard particles. 

Of course, within the present context, where the self--energy is independent
of the external momentum, there is no distinction between the
screening mass and the quasiparticle mass:  both are
solution to the same gap equation (\ref{GAP2}). But in QCD, these are distinct
quantities already at lowest order, and this difference will play an important
role in the construction of approximations in Sect.~\ref{sec:HTLqp}. 
Thus, although this may look at first
somewhat artificial here, we shall introduce different notations for 
the two masses.
We shall keep
the notation $m_D$ for the screening (Debye) mass, and denote the 
hard quasiparticle mass as
$m_\infty$: the latter characterizes the excitation spectrum at large 
momenta, typically of the
form $\varepsilon_k=\sqrt{k^2+m_\infty^2}$.  Thus
the mass
$m^2$ in the second order term of eq.~(\ref{Sdem}) should be read as 
$m_\infty^2$, while $m^3$
in the cubic term is $m_D^3$. That is, we rewrite eq.~(\ref{Sdem}) as
\beq \label{Sdem1}
{\cal S}(m)\simeq
\frac{2\pi^2 T^3}{45}-\frac{m_\infty^2
T}{12}+\frac{m_D^3}{12\pi}.
\eeq

\begin{figure}[t]
\centerline{\includegraphics[bb=70 200 540 540,width=6cm]{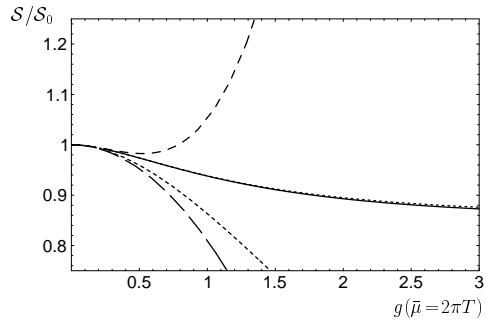}}
\caption{Comparison of the fully self-consistent 2-loop entropy
with the perturbative results to order $g^2$ and
$g^3$ (long and medium dashed lines, resp.). The short-dashed
line between the order $g^2$ result and the full line corresponds
to the ``HTL approximation'' where the 2-loop entropy is
evaluated with just the HTL mass; the short-dashed line
almost on top of the full line corresponds to the ``NLA approximation''
using (\protect\ref{minftself}).
}
\label{figscS}
\end{figure}

In perturbation theory, $m_D$ and $m_\infty$ coincide with each other 
up to order $g^3\,$,
but differ in order $g^4$ and higher.\footnote{Note that only
the corrections to $m_D$ can be calculated by means of
dimensional reduction. 
Dynamical masses in general require
the resummation of the nonstatic hard thermal loops
(for simple examples see Ref.~\cite{Kraemmer:1995az}).}
Within the 2--loop $\Phi$--derivable
approximation, they coincide if computed {\it exactly}, as already 
mentioned. But
  if the self--consistency is imposed only in the soft sector, the
corresponding approximations may a priori look different. We have 
seen earlier that
self-consistency in the soft sector is enough to obtain an excellent 
approximation for the
pressure, and we want to verify here the status of such an 
approximation for the entropy.
In fact we shall examine the accuracy of two non--perturbative approximations
to the entropy, that we shall later implement in the case of QCD. In 
both approximations,
the integral in the first line of eq.~(\ref{Sdem}) is evaluated exactly
(i.e., numerically), but for values of $m$ which are only approximate
solutions to the gap equation.

{\it i} ) In the `HTL approximation' we set $m=\hat m_D$.
The ensuing approximation for ${\cal S}$
includes the correct LO effect of the interactions, of order $g^2$,
but only 1/4 of the plasmon effect. In Fig. \ref{figscS}, 
the resulting approximation
is compared to the exact (i.e., fully self--consistent) result, and also to the
weak coupling expansion.

{\it ii} ) In the `next--to--leading approximation' (NLA), the mass $m_\infty$
of the {\it hard} particles is determined with NLO accuracy, so as to restore
the correct plasmon effect in  ${\cal S}$. That is,
\be\label{minftself}
m^2_\infty\,\longrightarrow  m^2_{\infty}  - (3/\pi)  g^2T m_D,\ee
where $m_D$ is
the self-consistent mass, given by the solution to the gap equation
(\ref{gapsoft}).
Of course, the mass $m_\infty$ thus obtained coincides
with the Debye mass $m_D$, but this `degeneracy' is specific to the
present scalar model, as we have already emphasized.

The quality of the NLA based on eq.~(\ref{minftself}) can be appreciated
from its comparison, in Fig. \ref{figscS},
with the ``exact'' (fully self-consistent)
result,
and also with the
HTL approximation alluded to before. Although the HTL approximation
is seen to deviate from the fully self-consistent result at large coupling,
it remains smaller than the ideal-gas result despite the
presence of a quarter of the plasmon effect, which, when
treated purely perturbatively, would result in a nonmonotonic
behaviour leading to an excess over the ideal-gas result
at large coupling.

The $\Phi$-derivable 2-loop approximation in fact becomes ``exact''
in the large-$N$ limit, where $N$ scalars are interacting in an
O($N$)-invariant manner according to $(\vec \phi^2)^2$. In this
limit the exact result is $\bar\mu$-independent, while all
approximations to it are renormalization scale dependent.
However, as can be seen in Fig. \ref{figscS}
the agreement is almost perfect for NLA when
$\bar\mu=2\pi T$ and varies only slightly when varying
this scale, which motivates us to adopt $\bar\mu=2\pi T$
as optimal renormalization point. However, in what follows we shall always
consider variations by a factor of 2 around this central
value to have an estimate of the inherent uncertainty
of our approximations.

\section{Quark--Gluon Plasma}

Before describing how to overcome the difficulties of perturbation
theory in the thermodynamics of QCD along the lines
sketched for the simple scalar models of the previous section, 
we shall briefly review the results of perturbation theory
obtained so far, which,
at least for vanishing chemical potential and high temperature,
have been recently pushed to the limits of calculability.

\subsection{Perturbative results}

In QCD with $N_f$ flavors of quarks, the thermodynamic pressure
at high temperature has been computed\cite{Arnold:1995eb,Zhai:1995ac} 
in perturbation theory through\break order~$g^5$:
\bea\label{Fpt}
P &=& {8\pi^2\045}T^4 \biggl\{
\left(1+{21\032}N_f\right)+
{-15\04}\left(1+{5\012}N_f\right){\alpha_s\0\pi}\nn&&\qquad\quad
+30\left[\left(1+{N_f\06}\right)\left({\alpha_s\0\pi}\right)\right]^{3/2}\nn
&&+\Bigl\{237.2+15.97N_f-0.413N_f^2+
{135\02}\left(1+{N_f\06}\right)\ln\left[{\alpha_s\0\pi}(1+{N_f\06})\right]\nn
&&\qquad\qquad-{165\08}\left(1+{5\012}N_f\right)\left(1-{2\033}N_f\right)
\ln{\bar\mu\02\pi T}
\Bigr\}\left({\alpha_s\0\pi}\right)^2\nn
&&+\left(1+{N_f\06}\right)^{1/2}\biggl[-799.2-21.96 N_f - 1.926 N_f^2\nn
&&\qquad\quad+{495\02}\left(1+{N_f\06}\right)\left(1-{2\033}N_f\right)
\ln{\bar\mu\02\pi T}\biggr]\left({\alpha_s\0\pi}\right)^{5/2}\nn&&\qquad\quad+
\mathcal O(\alpha_s^3\ln\alpha_s) \biggr\},
\eea
where $\alpha_s=g^2(\bar\mu)/(4\pi)$.
The coefficient of the $\alpha_s^3\ln\alpha_s$ term, the last
in the pressure at high $T$ and vanishing chemical potential
that is calculable in perturbation theory, has
recently been determined as\cite{Kajantie:2002wa}
\bea\label{Pg6}
P\Big|_{g^6\ln g}&=&
{8\pi^2\045}T^4 \biggl[1134.8+65.89 N_f+7.653 N_f^2\nn&&
-{1485\02}\left(1+{N_f\06}\right)\left(1-{2\033}N_f\right)
\ln{\bar\mu\02\pi T}\biggr]\left({\alpha_s\0\pi}\right)^{3}
\ln{1\0\alpha_s}\,.
\eea

Here $\bar\mu$ is the renormalization scale parameter of
the $\overline{{\rm MS}}$ scheme and $\alpha_s(\bar\mu)$ is the
running coupling whose beta function is
known to four-loop order\cite{vanRitbergen:1997va},
though we shall restrict ourselves
to its two-loop version in the following, because
lattice determinations of $T_c/\Lambda_{\overline{\hbox{\scriptsize MS}}}$
usually employ only the latter.

\begin{figure}
\centerline{\includegraphics[bb=70 200 540 540,width=6.4cm]{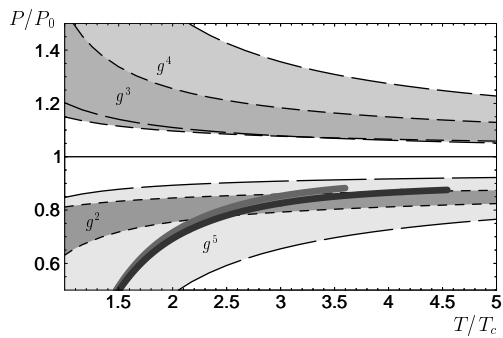}}
\caption{The poor convergence of thermal perturbation theory in
pure glue QCD. The various grey bands bounded by differently
dashed lines show the perturbative results
to order $g^2$, $g^3$, $g^4$, and $g^5$ with $\overline{\hbox{MS}}$
renormalization
point $\bar\mu$ varied between $\pi T$ and $4\pi T$. The thick
dark-grey line shows the lattice results from the Bielefeld group%
\protect\cite{Boyd:1996bx};
the lighter one behind that of a more recent lattice calculation
using an RG-improved action
from the CP-PACS collaboration\protect\cite{Okamoto:1999hi}.
\label{fig:qcd}}
\end{figure}

Evaluating the perturbative expression to order $g^2$, $g^3$, $g^4$, and $g^5$
gives results which do not show any sign of convergence, as depicted
in Fig.~\ref{fig:qcd}, but rather show increasing ambiguities due
to the dependence on the renormalization point, signalling a complete
loss of predictive power. (When the $g^6\ln g$-term of
Eq.~(\ref{Pg6}) is included with a suitably adjusted constant under the log,
the latter can at least be chosen such that
the lattice results are qualitatively reproduced\cite{Kajantie:2002wa}.)

However, the poor convergence behaviour of perturbation theory is
not very different from that we have observed above in the case
of simple scalar $\phi^4$ theory. In the following we shall
review the attempts to overcome this impasse by reorganizing perturbation
theory on the one hand by a variant of SPT and on the other by an approach
which is guided by $\Phi$-derivable approximations
for entropy and quark number density. In both
cases, the resulting expressions
go beyond strict perturbation theory in that higher-order terms
are kept unexpanded, and the hope is that strictly nonperturbative
contributions that appear in the pressure of hot QCD at order $g^6$
make up only a small, negligible part of what can be captured already
by an effective quasiparticle description as soon as the temperature
is a few times the transition temperature.

When compared with the perturbative results, it is remarkable
that the next-to-leading result to order $g^2$ fares rather
well at temperatures $\gtrsim 2T_c$, though the higher-order results
prove that perturbation theory is inconclusive. On the other
hand, simple quasiparticle models which describe the effective
gluonic degrees of freedom by $2N_g$ ($N_g=N^2-1$) scalar
degrees of freedom with asymptotic masses taken from a
HTL approximation can be used quite successfully to model
the lattice data by fitting the running coupling%
\cite{Peshier:1996ty,Levai:1997yx,Peshier:1999ww,Schneider:2001nf}.

\subsection{Lattice results}

The thermodynamic potential has been studied on 4-dimensional lattices
both for pure-glue QCD and QCD with a number of comparatively light
(but still massive) quarks by a number of groups.

In the case of pure-glue QCD, the temperature dependence of the
thermal pressure has been studied in great detail on lattices
with varying temporal extent up to $8\times 32^3$ for temperatures
up to about $4.5 T_c$. The continuum extrapolation using the
standard Wilson action of Ref.~\cite{Boyd:1996bx} is given in
Fig.~\ref{fig:qcd} by the darker-grey line with the width of the
line roughly representing the estimated error. Similar studies
have been performed using improved lattice actions%
\cite{Papa:1996an,
Okamoto:1999hi} and
the results obtained in Ref.~\cite{Okamoto:1999hi} using
a renormalization-group improved lattice action
are included in Fig.~\ref{fig:qcd} by the lighter-grey line.
The latter are
systematically higher by about $5\ldots 2\%$ for temperatures
$2\ldots 3.5T_c$ compared to the results of Ref.~\cite{Boyd:1996bx}.

As can be seen in Fig.~\ref{fig:qcd}, the pressure drops sharply
as the transition temperature is approached from above. Not shown
in the figure is that the pressure in fact drops to virtually zero
at and below the transition temperature. There the pressure is
exponentially suppressed by the rather large masses of glueballs.

In the presence of (light) quarks, lattice calculations are much
more difficult to perform and the continuum extrapolation is
correspondingly more involved and uncertain. However, through
the use of improved lattice action, considerable progress has
been made recently\cite{Karsch:2000ps,AliKhan:2001ek}.
When rescaled to the ideal-gas values, the results, as a function
of $T/T_c$ with the respective different values of $T_c$, the
pattern of the pressure of hot QCD with quarks is quite similar
to that of pure-glue QCD, and there is moreover a rather
weak dependence on the number of flavours if all of them are light.

\subsection{Dimensional reduction}

Dimensional reduction in hot QCD (at zero chemical potential)
leads to an effective three-dimensional
Lagrangian\cite{%
Appelquist:1981vg,Nadkarni:1983kb,Nadkarni:1988fh}
\be\label{LQCDdr}
\mathcal L_E=\2 \tr F_{ij}^2 + \tr [D_i,A_0]^2
+m_E^2 \tr A_0^2 + \2 \lambda_E (\tr A_0^2)^2 +\ldots
\ee
where the parameters are determined perturbatively by matching%
\cite{Braaten:1996jr,Kajantie:1997tt}. In lowest order one has a dimensionful
coupling $g_E^2 = g^2T$ and\cite{Nadkarni:1988fh}
\be\label{LQCDparam}
m_E^2=(1+N_f/6)g^2 T^2,\qquad
\lambda_E={9-N_f\012\pi^2}g^4T,
\ee
though $\lambda_E$ starts to contribute to the
pressure only at order $g^6$. At this order, however,
the self-interactions of the massless magnetostatic gluons
start to contribute, but these contributions are
inherently non-perturbative because the three-dimensional
theory for the zero modes $A_i(\vec x)$ is
a confining theory.\cite{Polyakov:1978vu,Linde:1980ts,Gross:1981br}

The thermal pressure of the 4-dimensional theory can be
decomposed into contributions from the hard modes $\sim T$,
calculable by standard perturbation theory, and soft contributions
governed by (\ref{LQCDdr}) which involves both perturbatively
calculable contributions up to order $g^5 T^4$ and the
nonperturbative ones starting at order $g^6 T^4$.

In Ref.~\cite{Braaten:1996jr}
the effective theory based on (\ref{LQCDdr}) has been used
to organize and reproduce the perturbative calculation of
the thermal pressure (\ref{Fpt}) of Refs.~\cite{Arnold:1995eb,Zhai:1995ac}.
This turns out to be particularly elegant when dimensional
regularization is used to provide both the UV and IR cutoffs
of the original and effective field theories.

To order $g^4$, the contribution of the hard modes can then be
written as\cite{Braaten:1996jr}
\bea\label{P3h}
P_{\rm hard}&=&{8\pi^2\045}T^4 \biggl\{\left(1+{21\032}N_f\right)+
{-15\04}\left(1+{5\012}N_f\right){\alpha_s\0\pi}\nn&&\quad
+\Bigl\{244.9+17.24N_f-0.415N_f^2
+{135}\left(1+{N_f\06}\right)\ln{\bar\mu\02\pi T}\nn
&&\qquad-{165\08}\left(1+{5\012}N_f\right)\left(1-{2\033}N_f\right)
\ln{\bar\mu\02\pi T}
\Bigr\}\left({\alpha_s\0\pi}\right)^2\biggr\}.
\eea
In the first logarithm the dimensional regularization scale $\bar\mu$ 
is associated with regularization in the infrared and thus has to
match a similar logarithm in the effective theory, whereas the second
logarithm is from UV and involves the first coefficient of the beta
function.

Indeed, calculating the pressure contribution of the soft sector
described by (\ref{LQCDdr}) in dimensional regularization
gives, to three-loop order (neglecting $\lambda_E$-contributions)
\bea\label{P3s}
P_{\rm soft}/T &=& {2\03\pi}m_E^3-{3\08\pi^2}\left(
4\ln{\bar\mu\02m_E}+3\right)g_E^2 m_E^2\nn&&
-{9\08\pi^3}\left({89\024}-{11\06}\ln2+{1\06}\pi^2\right)
g_E^4\,m_E^{\phantom4}.
\eea
All the contributions to the pressure involving odd powers of $g$
in (\ref{Fpt})
(as well as part of those involving even powers) are coming
from the soft sector. Inserting the leading-order value (\ref{LQCDparam})
for $m_E$ gives the QCD pressure up to and including order $g^4\ln g$;
to obtain all the terms to order $g^5$, next-to-leading order
corrections to the $m_E$-parameter have to be obtained
by a matching calculation as given in Ref.~\cite{Braaten:1996jr}.

In order to obtain the $g^6\ln g$-contribution (\ref{Pg6}) one
also needs $g_E^2$ to order $g^4$ (given in Ref.~\cite{Kajantie:1997tt})
and above all
the four-loop contribution of the effective theory (\ref{LQCDdr})
which has recently been calculated analytically as%
\cite{Kajantie:2002wa}
\bea\label{P4s}
P_{\rm soft}^{(4)}/T &=& N_g {(N g_E^2)^3 \0(4\pi)^4}
\left( \left[ {43\012}-{157\pi^2\0768} \right] \ln{\bar\mu\0g_E^2}
+ \left[ {43\04}-{491\pi^2\0768} \right] \ln{\bar\mu\0m_E}+c \right)\nn
\eea
up to a constant $c$ that
is strictly nonperturbative and needs to
be determined by three-dimensional lattice calculations. 
Such calculations have been carried out recently\cite{Kajantie:2000iz},
but they depend
on an as yet undetermined 4-loop matching coefficient. At the moment the
conclusion is that it is at least
not excluded that the lattice results
based on dimensional reduction can be matched to the full four-dimensional
results at temperatures of a few times the transition temperature.
For this reason, the most reliable results on the thermodynamics
of hot QCD (particularly for pure-glue\footnote{Inclusion of
fermions is particularly easy in the dimensional reduction
method, which makes a full three-dimensional prediction
clearly most desirable.} QCD) remain to date the
four-dimensional lattice data.

An interesting feature of the perturbative calculation when
organized through the dimensionally reduced effective theory
(\ref{LQCDdr}) is that the large scale dependences of strict
perturbation theory can be reduced when the effective parameters
are not subsequently expanded out when their perturbative results
are inserted in (\ref{P3s})\cite{Kajantie:2002wa,Blaizot:2003iq}.
What is more, considering successively the full two-loop and
full three-loop contributions from the soft sector, the results
do not exceed the ideal-gas result (as the strictly perturbative
results to order $g^3$ and $g^4$ do), and their scale dependence
diminishes by going from two to three-loop order.\cite{Blaizot:2003iq}

At three-loop order, it is in fact possible to eliminate the
scale dependence altogether by a principle of minimal sensitivity.
The result in fact agrees
remarkably well with the 4-d lattice results down to $\sim 2.5T_c$.
(This is also true when the four-loop logarithms are
included provided a suitable
value for the unknown constant
$c$ in (\ref{P4s}) is chosen.\cite{Kajantie:2002wa,Laine:2003ay})

While this goes only minimally beyond a strictly perturbative treatment,
it strongly suggests that perturbative QCD at high temperature, when
supplemented by appropriate resummation of soft physics, is capable of
quantitative predictions at temperatures of interest.


\section{Thermodynamics of HTL quasiparticles in QCD}
\label{sec:HTLqp}

\subsection{HTL-screened perturbation theory}

In the simple example of scalar $\phi^4$ theory, we have
seen that a promising strategy for improving the poor convergence
of perturbation theory may be screened perturbation theory
based on the reorganization of the Lagrangian according to
(\ref{effLagran}). The corresponding resummation
changes the UV structure of theory at any finite order
of perturbation theory, but removing the additional
UV divergences by some minimal subtraction does seem
to lead to results which avoid the bad behaviour of
conventional perturbation theory.

This method can be extended to QCD by employing the
gauge invariant HTL effective action%
\cite{Braaten:1992gm,Frenkel:1992ts,Blaizot:1994be}
instead of a simple
mass term. The HTL effective action is nonlocal and
involves nontrivial vertices in addition to self-energies, but it
is uniquely determined by only one mass parameter for gluons (and
a further one for quarks, if any)%
\cite{Andersen:1999fw,
Baier:1999db,%
Andersen:2002ey,Andersen:2003zk}.
This variant of screened perturbation theory has been termed
{\it HTL perturbation theory} (HTLPT)\footnote{We
would suggest however to spell this out as HTL-screened perturbation
theory to recall that this method transcends
the original HTL resummed perturbation theory\cite{Braaten:1990mz}.}
by their authors, but it should
be noticed that it is meant to transcend perturbative {\it HTL resummation},
where hard propagators and vertices with hard propagators attached
remain undressed\cite{Braaten:1990mz}.

In HTLPT the QCD Lagrangian is improved by writing
\be\label{LHTLPT}
\mathcal L_{\rm QCD}=\mathcal L_{\rm QCD}\Big|_{\alpha_s\to \delta \alpha_s}
+(1-\delta)\mathcal L_{\rm HTL}
\ee
with the standard HTL Lagrangian%
\cite{Braaten:1992gm,Frenkel:1992ts,Blaizot:1994be}
\be
\mathcal L_{\rm HTL}=-\2 m_D^2 \tr F_{\mu\alpha}\left\langle
v^\alpha v^\beta \0 (v\cdot D(A))^2 \right\rangle F^\mu{}_\beta
+iM^2 \bar\psi \gamma^\mu \left\langle v_\mu \0 v\cdot D(A)
\right\rangle \psi,
\ee
where $v^\mu=(1,\vec v)$ is a light-like four-vector and
$\langle\ldots\rangle$ represents the average over the
directions $\vec v$.
Since this only serves as gauge invariant generalization
of the thermal mass term for SPT, i.e.
a variationally improved perturbation theory, the
parameters $m_D$ and $M$ need no longer be restricted
to their leading-order HTL values (given in eqs.~(\ref{MD})
and (\ref{MF}) below).
Technically, HTLPT corresponds to using $\delta$ as an expansion
parameter that is set equal to 1 eventually.

In Ref.~\cite{Andersen:1999fw
}, HTLPT has been
used to calculate the pressure of QCD at one-loop level.
This corresponds to calculating the thermal pressure of
a modified free Lagrangian (\ref{LHTLPT}) with $\delta=0$.
If (but only if) the mass parameters $m_D$ and $M$ are identified
with their leading-order HTL values (\ref{MD})
and (\ref{MF}), this includes the plasmon term $\sim g^3$
of the pressure correctly, but over-includes the leading-order
interaction term $\sim g^2$ by a factor of 2. (In
Ref.~\cite{Andersen:1999fw
} the over-inclusion
factor was in fact larger because of an inconsistent usage
of dimensional regularization\cite{Blaizot:2000fc}. Corrected
numerical results at one-loop level are contained in
Refs.~\cite{Andersen:2002ey,Andersen:2003zk}.)

Starting at two-loop order, it becomes possible to determine
$m_D$ and $M$ by a variational principle. The two-loop pressure
in HTLPT is in fact almost prohibitively complicated because
of the necessity to include the non-local HTL vertices along
with the nontrivial HTL propagators.
Refs.~\cite{Andersen:2002ey,Andersen:2003zk} have simplified
this task by expanding the resulting sum-integrals in powers
of $m_D/T$ and $M/T$ such as to include all terms through order
$g^5$ if $m_D/T$ and $M/T$ are taken to be of order $g$.
(These mass expansions 
have been checked to be fairly accurate
in the applications of interest%
\cite{Andersen:2001ez}.) When
determining $m_D/T$ and $M/T$ through variational equations,
the two-loop HTLPT result is equivalent
with the perturbative result (\ref{Fpt}) through order
$g^4 \log(g)$, with deviations
setting in at order $g^4$.
The thermal mass parameter $m_D$ coincides with the
perturbative result for the Debye mass at leading order,
though not beyond, which is not too surprising since
it is just a variational parameter and not directly related
to the full electrostatic propagator; the quark mass parameter
$M$ when determined by a variational principle comes
out as $\sim g$ at weak coupling, however with
a coefficient that differs from
leading-order perturbation theory
and is in fact subtraction-scale dependent\cite{Andersen:2003zk}.

When evaluated numerically, the HTLPT results, like those of SPT in the
scalar case, turn out to
be a clear improvement over the conventional perturbative results
in that (at least for all $T\gtrsim 1.5 T_c$) both the one-loop and the
two-loop results are lower than the ideal-gas result while
including the plasmon effect $\sim g^3$ which in conventional
perturbation theory (see Fig.~\ref{fig:qcd}) leads to results much above
the ideal-gas value in contradiction with all lattice results.
Moreover, the one-loop and the
two-loop results do not deviate too strongly from each other so
that they seem to have satisfactory convergence. On the other
hand, they fall short of describing lattice results quantitatively
but remain too close to the ideal-gas value and account for
only about half of the deviation of the lattice results from
the latter.


\begin{figure}[t]
\centerline{\includegraphics[bb=70 200 540 555,width=6.6cm]{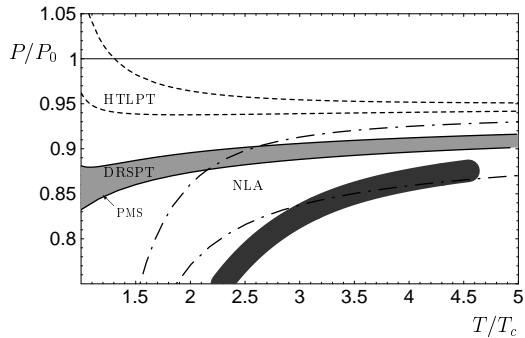}}
\caption{\label{fig:drspt}
The two-loop DRSPT result\protect\cite{Blaizot:2003iq} 
for pure-glue QCD (gray band) in comparison with
the two-loop HTLPT result of Ref.~\protect\cite{Andersen:2002ey} (dashed lines)
and the ``NLA''
result of Ref.~\protect\cite{Blaizot:2000fc} described in Sect.~\protect\ref{sec:PNLA}
(dash-dotted lines), all
with $\bar\mu$ varied around $2\pi T$ by a factor of 2.}
\end{figure}

A hint on the possible reason for this apparent
deficiency of HTLPT in the application
to the thermodynamics of QCD comes from a simpler variant
of SPT, namely when the latter is implemented only in the
dimensionally reduced sector (\ref{LQCDdr}) by rewriting
(\ref{LQCDdr}) trivially as
\be\label{EQCDSPT}
\mathcal L_{\rm EQCD}=
{1\04}F_{ij}^a F_{ij}^a + {1\02} (D_i A_0)^a (D_i A_0)^a
+{1\02}(m_E^2+\delta m^2) A_0^a A_0^a
-{1\02}\delta m^2 A_0^a A_0^a.
\ee
Here $m_E^2$ is determined by perturbative matching, whereas
the additional $\delta m^2$ is to be fixed by a gap equation
(i.e., a principle of minimal sensitivity). Like any SPT, this
dimensionally reduced SPT (DRSPT)
introduces additional UV singularities at finite loop order,
which should be minimally subtracted to allow comparison with HTLPT.

Like HTLPT, this formalism is gauge invariant, but very much
simpler and it avoids
the intermediate modification of hard modes (where ordinary perturbation
should be accurate anyway). 
In Ref.~\cite{Blaizot:2003iq} it has been recently
shown that the two-loop result 
of DRSPT
is much closer to the lattice results (reproduced in Fig.~\ref{fig:drspt}).
Because fermions are integrated out in DRSPT, this also avoids
the problem HTLPT appears to have with a gap equation for
the fermionic mass. (On the other hand and in contrast to HTLPT, 
DRSPT is clearly restricted
to static quantities.)

Unfortunately, at three-loop order DRSPT gives rise to a gap equation
for $\delta m^2$ which does not connect to perturbation theory any
longer so that it cannot be viewed as an improvement of perturbation
theory.\cite{Blaizot:2003iq}

\subsection{2PI formalism in gauge theories}
\label{sec:htlresummed}

In the following we shall describe a different approach in which
the 2PI formalism in terms of full propagators is used to
construct approximations which avoid the UV problems and
ambiguities of SPT and HTLPT. This turns out to
amount, at the two-loop level, to a description of
the thermodynamics of hot QCD in terms of weakly
interacting quasiparticles, whose nontrivial
spectral data serve to
incorporate (at least in leading and subleading orders) 
the strong interactions
of the elementary degrees of freedom.

In QCD, the thermodynamic potential is a functional of the full
gluon ($D$), quark ($S$), and Faddeev-Popov ghost ($D_{gh}$) propagators,
\bea\label{LWQCDwgh}
\b \Omega[D,S,D_{gh}]&=&\2 \Tr \log D^{-1} - \Tr \log S^{-1} - \Tr \log D^{-1}_{gh}
\nn&&
- \2 \Tr \Pi D + \Tr \Sigma S + \Tr \Pi_{gh} D_{gh} + \Phi[D,S,D_{gh}],
\eea
where Tr now includes traces over color indices, and also over
Lorentz and spinor indices when applicable. The self-energies
for gluons, quarks and ghosts are denoted by
$\Pi$, $\Sigma$ and $\Pi_{gh}$, respectively.

\begin{figure}
\epsfxsize=6cm
\centerline{\epsfbox[40 350 505 470]{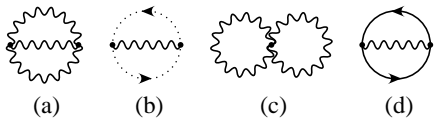}}
\vskip0.5cm
\caption{Diagrams for $\Phi$ at 2-loop order in QCD.
Wiggly, plain, and dotted lines refer respectively to
gluons, quarks, and ghosts.
\label{figphiqcd}}
\end{figure}

In the following we shall consider a two-loop $\Phi$-derivable
approximation which amounts to keeping the diagrams displayed in
Fig.~\ref{figphiqcd}. This again leads to the important simplification
that the entropy is given by a comparatively simple expression
\bea
\label{S2loop}
{\cal S}&=&-\tr \int{d^4k\0(2\pi)^4}{\6n(\omega)\0\6T} \left[ \Im
\log D^{-1}-\Im \Pi \Re D \right] \nn
&&-2\tr \int{d^4k\0(2\pi)^4}{\6f(\omega)\0\6T} \left[ \Im
\log S^{-1}-\Im \Sigma \Re S \right],
\eea
with vanishing interaction term $\mathcal S'$ (cf.~(\ref{SP0})), 
and this also holds
true for fermion number densities:
\bea
\label{N2loop}
{\cal N}&=&-2\tr \int{d^4k\0(2\pi)^4}{\6f(\omega)\0\6\mu} \left[ \Im
\log S^{-1}-\Im \Sigma \Re S \right],
\eea
where the completely analogous interaction term $\mathcal N'$
is equally absent at the two-loop level.

While $\Phi$-derivable
approximations are symmetry preserving as concerns global
symmetries\cite{Baym:1962}, local gauge invariance
is not respected, so this approximation introduces gauge dependences
as long as vertices are not dressed in a self-consistent
manner in line with the self-energies\footnote{On a formal
level this has been worked out in Ref.~\cite{Freedman:1977xs}.}.
These gauge dependences are actually parametrically suppressed
at the stationary point%
\cite{Arrizabalaga:2002hn}, where they enter only at twice the
order of the truncation. However,
in what follows we shall
construct further approximations based on the
gauge invariant hard thermal loops, and this will
entail manifest gauge independence in our applications.
In these approximations, where the self-consistency of the
two-loop $\Phi$-derivable scheme is only approximately realized,
but such that the deviations from full self-consistency are
of the order of three-loop corrections,
only the two physical polarizations of
the gluons contribute and the Faddeev-Popov ghosts just cancel
the unphysical gauge degrees of freedom of the former so that
they can be simply dropped.

\subsection{
Approximately self-consistent
entropy and quark density}

As discussed above, conventional thermal perturbation theory
exhibits very poor apparent convergence as soon as collective
phenomena such as the plasmon effect are included, which
involve odd powers in the coupling. The self-consistent
entropy and density functionals at two-loop order
also contain the plasmon effect, but together with
higher-order terms that ordinarily would have been discarded
by a strictly perturbative expansion in $g$ because otherwise
the renormalization programme could not have been carried out.
By contrast, the two-loop entropy and density functionals
(\ref{S2loop}) and (\ref{N2loop}) are UV finite expressions
and can be evaluated in a nonperturbative manner using
the high-temperature/density approximation of the gluon
and quark propagators, which are finite at leading and
next-to-leading order in HTL perturbation theory.

\subsubsection{HTL approximation}

As a first approximation we consider the two-loop
entropy and density functionals evaluated completely using
HTL (HDL) propagators. From the discussion of the scalar model
we anticipate that this includes correctly the leading-order
interaction contribution $\propto g^2$ in the entropy (in contrast
to the pressure).\footnote{A different procedure for
including the full HTL propagators has been proposed by
Peshier\cite{Peshier:2000hx}, who considers a
(somewhat ad-hoc) modification
of the $\Phi$ functional such as to have correct leading-order
contributions and UV finiteness
when full propagators are approximated by HTL ones. This turns out
to be equivalent to our results when expanded out to order $g^3$.}

But unlike the scalar $\phi^4$ model, where the HTL self-energy
is a simple constant mass,
cf. (\ref{Sigma2}), the HTL self-energies for gluons and quarks contain
two non-local structure functions each.
The gluon HTL self-energy contains transverse and longitudinal
components given by
\bea
\label{PiTL}
\5\Pi_L(\omega,k)&=&
\5m_D^2\left[1-{\omega\02k}\log{\omega+k\0\omega-k}\right],\nn
\5\Pi_T(\omega,k)&=&\frac{1}{2}\left[\5m_D^2+\,\frac{\omega^2 - k^2}{k^2}\,
\5\Pi_L\right],
\eea
where
\be\label{MD}
\hat m^2_D=(2N+N_f)\,\frac{g^2T^2}{6}\,+\,N_f\,{g^2\mu^2\02\pi^2}
\ee
is the (leading-order) Debye screening mass.\footnote{For simplicity
we shall write out our formulae for only one value of the chemical
potential; the generalization to several different chemical potentials is
straightforward. }
As is well known\cite{Kalashnikov:1980cy,Weldon:1982aq}, the gluon propagator
dressed by these self-energies
has quasiparticle poles at $\omega_{T,L}(k)$
with momentum-dependent effective masses
and Landau damping cuts for $|\omega|<k$. When $k\gg \5m_D$, the
pole corresponding to the collective longitudinal
excitation has exponentially vanishing residue\cite{Pisarski:1989cs}, whereas
transverse excitations tend to $\omega\to\sqrt{k^2+m_\infty^2}$, where
the asymptotic mass is given by
\be\label{mas}
m_\infty^2=\5\Pi_T(k,k)=\2 \5m_D^2\,.
\ee

The (massless) quark propagator at finite temperature or density is split into
two separate branches of opposite ratio of chirality over helicity
with propagators
$
\Delta_\pm=[-\omega+k \pm \Sigma_\pm]^{-1}
$.
In the HTL approximation, the respective self-energies read
\cite{Klimov:1981ka}:
\be\label{SIGHTL}
\hat\Sigma_\pm(\omega,k)\,=\,{\5M^2\0k}\,\left(1\,-\,
\frac{\omega\mp k}{2k}\,\log\,\frac{\omega + k}{\omega - k}
\right),\ee
where ($C_f=(N^2-1)/2N$)
\be\label{MF}
\5M^2\,=\,{g^2 C_f\08}\left(T^2+{\mu^2\0\pi^2}\right).\ee

The dressed fermion propagators have quasiparticle poles
at $\omega_\pm(k)$ and Landau damping cuts.
At large $k$ and positive frequency
$\omega_+\to\sqrt{k^2+M_\infty^2}$
with asymptotic mass $M_\infty^2=2\5M^2$;
the additional branch $\omega_-$
which has no analogue at zero temperature
and density has exponentially vanishing residue%
\cite{Pisarski:1989wb,Blaizot:1993bb}.

Although the various HTL self-energies constitute a leading-order
result only under the condition of {\em soft} momenta and frequencies,
the result for the asymptotic masses applies also for {\em hard} momenta%
\cite{Kraemmer:1990dr,Flechsig:1996ju}.

Using these expressions in the entropy formula (\ref{S2loop})
this takes care of all
contributions of order $g^2$, but only part ($\le 1/4$) of
the plasmon term $\sim g^3$. However, it also contains
infinitely many higher-order terms which despite
being incomplete may help to get rid of the pathological
behaviour of the perturbation series truncated at
low orders in $g$. Among such higher-order contributions
is for instance a $g^4$ contribution reading (for pure glue)
\be\label{S4HTL}
{\cal S}_{HTL}^{(4)}=-N_g{\hat m_D^4 \0 16\pi^2 T}
\(\log{T\0\hat m_D}+1.55\ldots \)
\ee
involving a $g^4\log(1/g)$-term, whose coefficient in pure-glue QCD
is 1/12 of that of the complete perturbative result.\footnote{The
correct coefficient will be restored by $O(g^4\log(1/g)T^2)$ corrections
to $m_\infty^2$, whereas the constant under the logarithm also
receives three-loop contributions.}
By contrast, simple massive quasiparticle models such as
those used in Refs.~\cite{Peshier:1996ty,Levai:1997yx}
do not have a $g^4\log(1/g)$-term in the entropy at all.


\begin{figure}[htb]
\centerline{\includegraphics[bb=70 200 540 560,width=6.4cm]{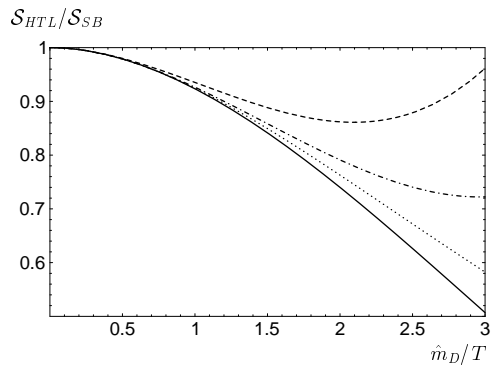}}
\caption{HTL entropy per gluonic degree of freedom normalized to
its Stefan-Boltzmann value as a function of the Debye mass
$\hat m_D(T,\mu)/T$.
The full line gives the complete numerical
result; the dashed line corresponds
to the perturbative result to order $(\hat m_D/T)^3\sim g^3$. The
dotted line gives the entropy for scalar degrees of freedom with
momentum-independent mass $m=m_\infty=\hat m_D/\sqrt2$; its
perturbative approximant is given by the dash-dotted line.}
\label{figSS0g}
\end{figure}

In Fig.~\ref{figSS0g}, the numerical result for ${\cal S}_{HTL}/{\cal S}_{SB}$
in the case of pure glue is given as a function of $\hat m_D/T$, which
is the only independent parameter. The HTL result is depicted by the
full line and is found to be a monotonically decreasing function
of $\hat m_D/T$. If this were expanded in powers of $\hat m_D/T$
and truncated beyond $(\hat m_D/T)^3\sim g^3$
(dashed line in Fig.~\ref{figSS0g}), this property would
have been lost at $\hat m_D/(2\pi T)\approx 1/3$, where one might
still expect a sufficiently clear separation of hard and soft
scales which is a prerequisite of the HTL approximation.

The numerical result for ${\cal S}_{HTL}$ is in fact very close
to a simple massive
quasiparticle model with entropy $2N_g {\cal S}_{SB}(m_\infty)$,
represented by the dotted line in Fig.~\ref{figSS0g}.
This is due to large cancellations between the longitudinal and
transverse contributions to the HTL entropy. However,
quite remarkably
one has ${\cal S}_{HTL}<2N_g {\cal S}_{SB}(m_\infty)$ even
though the latter contains 30\% less of the plasmon term $\sim g^3$, which
would be expected to work in the opposite direction.
This is further emphasized by comparison with its perturbative
approximation given by the dash-dotted line.

\begin{figure}[htb]
\centerline{\includegraphics[bb=70 200 540 560,width=6.4cm]{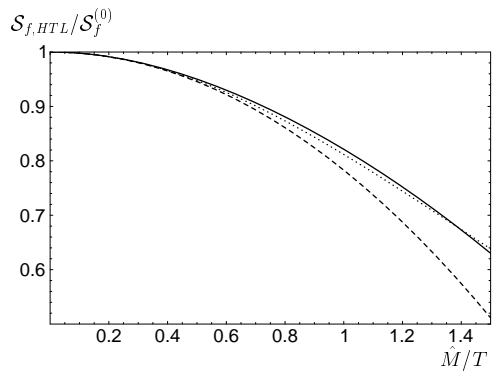}}
\caption{HTL entropy per quark degree of freedom at $\mu=0$ normalized
to its free value (solid line), the corresponding perturbative
order-$g^2$ result (dashed line), and the entropy of
a quark with constant mass $M_\infty$ (dotted line).}
\label{figSS0q}
\end{figure}

In Fig.~\ref{figSS0q} the HTL entropy of 1 quark degree of freedom
at zero chemical potential is displayed as a function of
the fermionic plasma frequency $\hat M$. Like in the
pure-glue case, this contains also an (incomplete) $g^4\log(1/g)$
contributions that is not present in simpler quasiparticle models:
\be
{\cal S}_{f,HTL}^{(4)}=NN_f{\hat M^4\0\pi^2 T}\( \log{T\0\hat M}+0.22\ldots \)
\ee
${\cal S}_{f,HTL}$ does not contain any contributions to the
plasmon term $\sim g^3$, which entirely come from NLO corrections
to $M_\infty$. There is therefore rather little deviation of the
full numerical result for the fermionic contribution (solid line) from the
perturbative one truncated above order $g^2$ (dashed line).

Compared to the entropy of a simple massive fermionic
quasiparticle model ${\cal S}_{f}^{(0)}(M_\infty)$ (dotted line
in Fig.~\ref{figSS0q}), one finds extremely good numerical agreement,
which again takes place only after adding up all the quasiparticle
($(+)$ and $(-)$) and Landau-damping contributions, however.

In the case of zero temperature but high chemical potential,
the quantity of interest is the quark density. This does
not contain any plasmon term $\sim g^3$, but rather
$\sim g^4 \log(1/g)$. The HDL approximation does contain
some though not all of this term:
\be
{\cal N}_{HDL}^{(4)}= NN_f{\hat M^4\0\pi^2 \mu}\( \log{\mu\0\hat 
M}+0.35\ldots \)
\ee
Order-$g^2 \log g$
contributions to the asymptotic masses of the quark and gluon quasiparticles,
still within the framework of the 2-loop $\Phi$-derivable approximation,
are responsible for the
remaining contribution to the coefficient of the $g^4\log g$-term,
while the coefficient under the logarithm also receives 3-loop contributions.

\begin{figure}[htb]
\centerline{\includegraphics[bb=70 200 540 560,width=6.5cm]{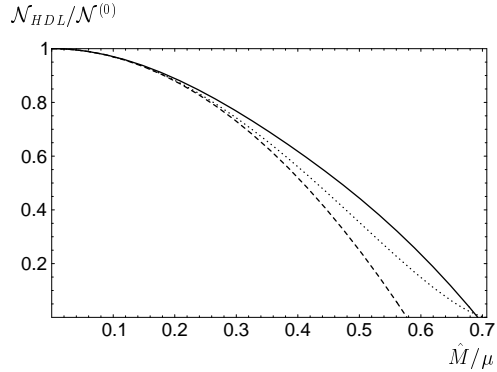}}
\caption{HDL quark density at $T=0$ normalized
to its free value (solid line), the corresponding perturbative
order-$g^2$ result (dashed line), and the free quark density of
a quark with constant mass $M_\infty$ (dotted line).}
\label{figNN0}
\end{figure}

In Fig.~\ref{figNN0} the numerical result for ${\cal N}_{HDL}$ is given
as a function of $\hat M/\mu$ (full line) and compared to that of
a simple quasiparticle model
\be\label{NT0}
{\cal N}_0(M_\infty)\Big|_{T=0}={1\03\pi^2}(\mu^2-M_\infty^2)^{3\02}
\theta(\mu-M_\infty)
\ee
as well as to a perturbative approximation truncated beyond $(\hat M/\mu)^2\sim
g^2$. Remarkably, the full HDL result drops to zero at almost the
same value as the simple quasiparticle model. However, the former
becomes negative thereafter, showing that the HDL approximation
is breaking down at $M_\infty/\mu > 1$ at the latest.

\subsubsection{Next-to-leading approximations}

The plasmon term $\sim g^3$ at high temperatures
$T\gg \hat m_D$ becomes complete only after inclusion
of the next-to-leading correction to the asymptotic thermal
masses $m_\infty$ and $M_\infty$. These are determined
in standard HTL perturbation theory through
\be\label{dPias}
\begin{array}{l}
\delta m_\infty^2(k)=\Re \delta\Pi_T(\omega=k) \\
=\Re(\begin{picture}(0,0)(0,0)
\put(25,0){\small +}
\put(56,0){\small +}
\put(104,0){\small +}
\put(157,0){$|_{\omega=k}$}
\end{picture}
\!\!\includegraphics[bb=145 430 500 475,width=5.5cm]{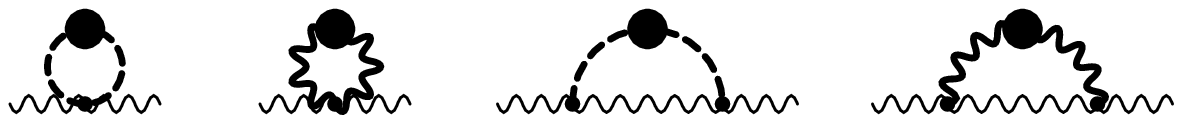})
\end{array}
\ee
where thick dashed and wiggly lines with a blob represent
HTL propagators for longitudinal and transverse polarizations, respectively.
Similarly,
\be\label{dSigmaas}
\begin{array}{l}
{1\02k}\delta M_\infty^2(k)=\delta\Sigma_+(\omega=k) \nonumber\\
=\Re(\begin{picture}(0,0)(0,0)
\put(42,0){\small +}
\end{picture}
\includegraphics[bb=75 430 285 475,width=3.2cm]{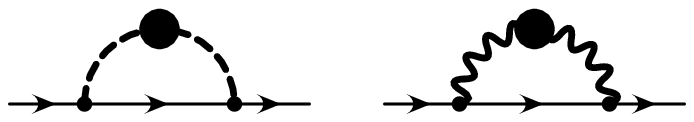})
|_{\omega=k}\;.
\end{array}
\ee

The explicit proof that these contributions indeed restore the
correct plasmon term is given in Ref.~\cite{Blaizot:2000fc}.

These corrections to the asymptotic thermal masses are, in contrast
to the latter, nontrivial functions of the momentum, which can
be evaluated only numerically. However, as far as the generation
of the plasmon term is concerned, these functions contribute
in the averaged form
\be\label{deltamasav}
\bar\delta m_\infty^2={\int dk\,k\,n_{\rm BE}'(k) \Re \delta\Pi_T(\omega=k)
\0 \int dk\,k\,n_{\rm BE}'(k)}
\ee
and similarly
\be\label{deltaMasav}
\bar\delta M_\infty^2={\int dk\,k\,n_{\rm FD}'(k) \Re
2k \delta\Sigma_+(\omega=k)
\0 \int dk\,k\,n_{\rm FD}'(k)}\;.
\ee
which can be determined analytically%
\cite{Blaizot:1999ap}, leading to the remarkably
simple expressions
\bea
\label{deltamas}
\bar\delta m_\infty^2=-{1\02\pi}g^2NT\hat m_D,&&\\
\label{deltaMas}
\bar\delta M_\infty^2=-{1\02\pi}g^2C_fT\hat m_D,&&\!\!\!\! C_f=N_g/(2N),\;
\eea

These corrections only pertain to the hard excitations;
in contrast to the scalar case, 
in gauge theories the one-loop
corrections to the various thermal masses of soft excitations
are known to differ substantially from (\ref{deltamas})
or (\ref{deltaMas}). For instance, the gluonic
plasma frequency at $k=0$ reads\cite{Schulz:1994gf}
$\delta m^2_{pl.}/\hat m^2_{pl.}\approx
-0.18\sqrt{N}g$, which is only about a third of
$\bar\delta m_\infty^2/m_\infty^2$; the NLO correction to the
nonabelian Debye mass on the other hand is even positive
for small coupling and moreover logarithmically enhanced%
\cite{Rebhan:1993az}
$\delta m^2_D/\hat m^2_D = +\sqrt{3N}/(2\pi) \times g \log(1/g)$.

For an estimate of the effects of a proper incorporation of the
next-to-leading order corrections we have therefore proposed
to include the latter only for hard excitations and to
define our next-to-leading approximation (for gluons) through
\be
{\cal S}_{NLA}={\cal S}_{HTL}\Big|_{\rm soft}+
{\cal S}_{\bar m_\infty^2}\Big|_{\rm hard},
\ee
where $\bar m_\infty^2$ includes (\ref{deltamas}).
To separate soft ($k\sim \hat m_D$) and hard ($k \sim 2\pi T$) momentum
scales, we introduce the intermediate scale
$\Lambda=\sqrt{2\pi T\hat m_D c_\Lambda}$
and consider a variation of $c_\Lambda=\2\ldots 2$ as part of
our theoretical uncertainty.

Another crucial issue concerns the definition of the corrected
asymptotic mass $\bar m_\infty$. For the range of coupling constants
of interest ($g> 1$), the correction $ |\bar\delta m_\infty^2|$
is greater than the LO value $m_\infty^2$, leading to tachyonic
masses if included in a strictly perturbative manner.

In the simple scalar model of Sect.~\ref{secsimplemodel} we have
seen that the requirement of self-consistency for the soft modes
leads to a quadratic gap equation for the asymptotic thermal
mass, eq.~(\ref{minftself}).

In QCD, where the (non-local) gap equations
are much too complicated to
be attacked directly, we simply consider perturbatively equivalent
expressions for the corrected $\bar m_\infty$ which are monotonic
functions in $g$. Besides the solution to a quadratic equation
analogous to (\ref{minftself}), namely
\be\label{minftyqgap}
\bar m_\infty^2=
{g^2(N+N_f/2)T^2\06}
-{g^2NT\0\sqrt2\pi}\bar m_\infty,
\ee
we have tried the
simplest Pad\'e approximant 
(degree [2,1]), which
also gives a greatly improved approximation to the solution of
the gap equation
of the scalar (large $N$) model. In QCD, our final results do not depend
too much on whether we use the Pad\'e approximant%
\cite{Blaizot:1999ip,Blaizot:1999ap}
or a quadratic gap equation%
\cite{Blaizot:2000fc}.

The main uncertainty rather comes from the choice of the
renormalization scale which determines the magnitude of
the running strong coupling constant (chosen to satisfy
the renormalization group equation to 2-loop order in the following).

\begin{figure}[tb]
\centerline{\includegraphics[bb=70 200 540 560,width=6.5cm]{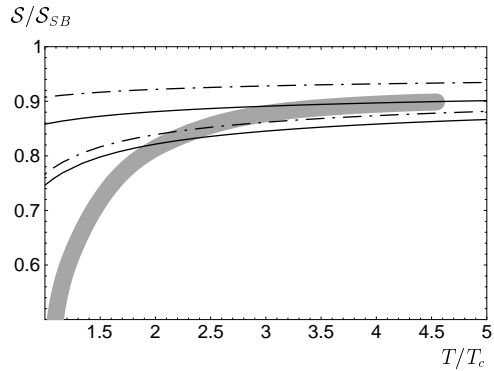}}
\caption{Comparison of the lattice data for
the entropy of pure-glue SU(3) gauge theory of Ref.~\protect\cite{Boyd:1996bx}
(gray band) with the range of ${\cal S}_{HTL}$ (solid lines)
and ${\cal S}_{NLA}$ (dash-dotted lines) for $\bar\mu=
\pi T\ldots 4\pi T$ and $c_\Lambda= 1/2 \ldots 2$.
}
\label{figSg}
\end{figure}

In Fig.~\ref{figSg}, the numerical results for the HTL entropy
and the NLA one are given as a function of $T/T_c$ with $T_c$
chosen as $T_c=1.14\Lambda_{\overline{\mathrm MS}}$. The full lines
show the range of results for
${\cal S}_{HTL}$ when the renormalization scale $\bar\mu$
is varied from $\pi T$ to $4\pi T$; the dash-dotted lines mark
the corresponding results for ${\cal S}_{NLA}$ with the
additional variation of $c_\Lambda$ from $1/2$ to 2. The
dark-gray band are lattice data from Ref.~\cite{Boyd:1996bx}
(the more recent results from Ref.~\cite{Okamoto:1999hi} are consistent
with the former within error bars and centered around the upper
boundary of the gray band for $T\approx 3T_c$ and somewhat flatter
around $2T_c$).
Evidently, there is very good agreement for $T>2.5T_c$.

When $N_f>0$ massless quarks are included,
the next-to-leading corrections to the asymptotic fermionic
masses (\ref{deltaMas}) have the same problem as the bosonic
ones that for large coupling the asymptotic masses may turn
tachyonic.
In Ref.~\cite{Blaizot:2000fc} they have therefore been treated
like the bosonic asymptotic masses, by postulating the simplest
quadratic gap equation which is perturbatively equivalent to
(\ref{deltaMas}). However, there is an obvious asymmetry in that
only gluonic quantities appear on the right-hand side of
(\ref{deltamas}) and (\ref{deltaMas}). Indeed, treating them
in the same manner is not consistent in the limit of
$N_f\to\infty$, which has been solved recently in
Refs.~\cite{Moore:2002md
}. To be consistent with
this limit, the fermionic gap equation has to remain linear
as it is in perturbation theory.
We therefore restrict our requirement of a perturbative equivalent
gap equation with monotonic behaviour only to the bosonic (gluonic)
sector. For the fermions we proceed by treating the right-hand
sides of eqs.~(\ref{deltamas}) and (\ref{deltaMas}) in the
same manner, which amounts to\cite{Rebhan:2003fj}
\be\label{newMgap}
\bar M_\infty^2={g^2 N_f C_f T^2\04}
- {1\0\sqrt2 \pi}g^2 C_f T \bar m_\infty.
\ee
There is then no negative feedback from the fermion mass itself, it
only inherits higher-order terms from the solution to $\bar m_\infty$.
This does not avoid the appearance of tachyonic masses when
both $N_f$ and $g$ are large, but it turns out that for $N_f\le 3$
one still has a monotonic behaviour of $M_\infty$ as a function
of $g$. In fact, in the case $N_f=3$ the solutions to the correspondingly
corrected gap equations happen to coincide precisely with those of
the (structurally rather different) decoupled quadratic equations
adopted in Ref.~\cite{Blaizot:2000fc}. As a result, the
changes on the results of Ref.~\cite{Blaizot:2000fc}
are rather minimal for phenomenologically interesting
values of $N_f$. (In the limit $N_f\to\infty$, there are more
important effects, such that the picture of stable quasi-particles
seems to have to be abandoned above some large coupling, but
up to this point the quantitative agreement with the exact result
is quite good\cite{Rebhan:2003fj}.)

\begin{figure}[tb]
\centerline{\includegraphics[bb=70 200 540 560,width=6.5cm]{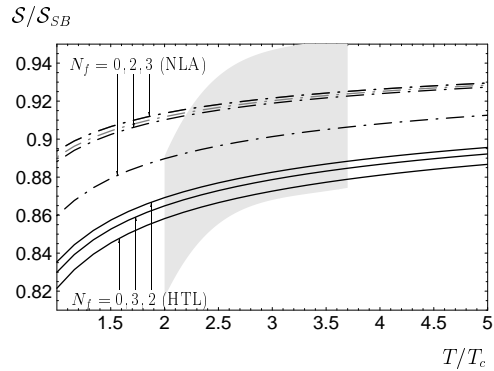}}
\caption{Comparison of the HTL entropy (solid lines) and the NLA
estimate (dash-dotted lines)
for $N_f=0,2,3$ with
the estimated continuum extrapolation of $N_f=2$ lattice data
of Ref.~\protect\cite{Karsch:1999vy}.
(The gray dash-dotted line corresponds to using a quadratic
fermionic gap equation, which only has an effect for NLA $N_f=2$,
which is minimally higher than the result obtained with a
linear one required for consistency with the large-$N_f$
limit\protect\cite{Rebhan:2003fj}).}
\label{figSNf023f}
\end{figure}

In Fig.~\ref{figSNf023f}, the (updated) results for the entropy with
$N_f \le 3$ massless quarks are
compared with a recent estimate\cite{Karsch:1999vy}
of the continuum limit of
lattice results for $N_f=2$ (gray band),
but now with ${\cal S}_{HTL}$ and ${\cal S}_{NLA}$
evaluated for the central choice
of $\bar\mu=2\pi T$ and $c_\Lambda=1$ (with unchanged
$T_c/\Lambda_{\overline{\mathrm MS}}$). When $N_f$ is increased,
there are competing effects of larger thermal masses versus
slower running of $\alpha_s$, which result into a rather
weak dependence of our results on $N_f$ as a function
of $T/\Lambda_{\overline{\mathrm MS}}$ as it is in Fig.~\ref{figSNf023f}.
This seems to be consistent with the very small $N_f$ dependence
observed on the lattice (when the thermodynamic potentials are normalized by
their ideal-gas limit and expressed as a function of $T/T_c$).

It should be understood, however, that
the simple quadratic gap equation (\ref{minftyqgap}) and
its extension to the fermionic sector, eq. (\ref{newMgap}), are
at present only an estimate
of how the next-to-leading order corrections
(\ref{dPias}) and (\ref{dSigmaas}) in actuality modify the spectral
properties of the hard quasiparticles. Moreover, the former
are included so far in the averaged form (\ref{deltamasav})
and (\ref{deltaMasav}). A possible
refinement is to include the non-trivial momentum dependence
of (\ref{dPias}) and (\ref{dSigmaas}) which we expect to
render a separation scale between hard and soft quasiparticles
superfluous because the corrections
(\ref{dPias}) and (\ref{dSigmaas})
appear to grow with momentum so that their incompleteness
at soft momenta may be negligible. When doing so we also hope to
be able to come to a better understanding of the mechanism
which provides the nonlinear feedback preventing
tachyonic masses from appearing.

\subsection{Pressure}
\label{sec:PNLA}

In the above, we have concentrated on the entropy (and quark number
density) as the thermodynamic potential which is most suitable
for a (HTL) quasiparticle description.

Were it not for the trace anomaly\cite{Drummond:1999si}
$\langle T^\mu_\mu \rangle = {\cal E}-3P \not=0$
the corresponding
pressure would be simply given by $P=(T{\cal S} + \mu{\cal N})/4$
(in the case of massless quarks).
The correct relation is instead provided by an integration such as
\be
P(T,\mu)=\int_{T_1}^T {\cal S}(T',\mu)\,dT'+P(T_1,\mu)\,.
\ee

Up to a single integration constant, entropy and density evidently
contain all the relevant information. This integration constant, however,
represents strictly nonperturbative information, because it amounts
to fixing the bag constant. Since QCD is asymptotically free, but
confining at low temperature, this constant cannot be fixed by
requiring that the thermal pressure as obtained from entropy and
density vanishes at zero temperature. It is also not possible to
determine $P(T_1)$ from the
requirement that both $p(\a_s)=P(T)/P_0(T)$ and $s(\a_s)=S(T)/S_0(T)$
approach 1 in the limit of $\a_s\to 0$, because the
differential
equation $p(\a_s)+\4\b(\a_s)p'(\a_s)=s(\a_s)$ with
$\b(\a_s)=-\b_0 \a_s^2-\b_1 \a_s^3-\dots$
has as homogeneous solutions
$$p(\a_s)_{\rm hom.}=C\exp\left\{-{1\0\a_s}[4\b_0^{-1}+O(\a_s)]\right\},$$
and this vanishes at $\a_s=0$
together with all its derivatives.

Instead we can fix the integration constant by comparison with lattice
data for the pressure at $\mu=0$. A simple possibility for our purposes is to
take $T_1=T_c$ and $P(T_c,0)=0$. For $T$ substantially larger than
$T_c$ this does not influence much the results derived from entropy
and density, since in the ratio $P/P_0$ the (bag) constant term
decays like $T^{-4}$.

The result obtained from integrating
the entropy in the case of pure-glue QCD
with boundary condition $P(T_c)=0$ is shown
in Fig.~\ref{fig:drspt} and compared with available lattice
data as well as the results obtained 
recently in 2-loop
HTLPT by Andersen {\it et al.}\cite{Andersen:2002ey}
and in a dimensionally reduced variant of SPT\cite{Blaizot:2003iq}.

\subsubsection{Finite chemical potential}

For finite chemical potential, the entropy and quark density
expressions satisfy the Maxwell relations
\be\label{Maxrel}
{\6 {\cal S}\0\6\mu}\Big|_T={\6 {\cal N}\0\6T}\Big|_\mu
\ee
only to the same order in $g$ to which they are accurate
when expanded out perturbatively.

In simple quasiparticle models\cite{Peshier:1996ty,Levai:1997yx}
one therefore has to allow for
a bag function (instead of a bag constant) to restore
thermodynamic consistency. This can in fact be
utilized for matching lattice results and mapping them
in a phenomenological manner to
nonvanishing chemical potential\cite{Peshier:1999ww},
where recently significant progress
has been made on the lattice side%
\cite{Fodor:2001au
}. This includes by now concrete results on the thermodynamical
potentials\cite{Fodor:2002km}, which have been found to
agree well with the simple quasiparticle models%
\cite{Szabo:2003kg}.

A similar programme has been carried through with the HTL expressions
for entropy and density in place of those of the simpler quasiparticle
models with constant masses, and the results are, like those
of the simpler models, encouraging\cite{Romatschke:2002pb}.
An extension which includes next-to-leading corrections to
the asymptotic thermal masses is work in progress.

\subsection{Quark number susceptibilities}

Quark number susceptibilities (QNS) are defined as the second derivatives
of the pressure with respect to the chemical potentials of the
different quark flavors:
\be\label{def-chi}
\chi_{ij}\equiv {\6 {\mathcal N}_i \0 \6 \mu_j}
  = {\6^2 {P} \0 \6 \mu_i\6 \mu_j}=
\chi_{ji}
\ee
where $i,\,j$ are flavor indices.

These quantities are of direct interest for studying the
transition to the quark-gluon plasma phase as they can be
related to fluctuations in conserved charges which can
discriminate against a purely hadronic phase%
\cite{Asakawa:2000wh
}.

On the theoretical side, there are new
results on QNS from lattice gauge theory%
\cite{Gavai:2001ie,Gavai:2002kq,Gavai:2002jt,Bernard:2002yd},
extending them to higher $T/T_c$ which can be compared against
conventional perturbative results and those resulting from resummation
schemes aiming at overcoming the poor convergence of the former.

\subsubsection{Diagonal quark number susceptibilities}

The diagonal QNS, which for degenerate quark flavors are all
equal and denoted simply by $\chi$ in what follows, are found
to increase sharply at the deconfinement phase transition toward
a large percentage of the ideal gas value $\chi_0=NT^2/3$ (for
SU($N$) and massless quarks).

In perturbation theory, the series expansion of $\chi/\chi_0$
is now known\cite{Vuorinen:2002ue} through order $g^6\ln(1/g)$
and the coefficient under the log at order $g^6$ is in fact
computable in perturbation theory, in contrast to the one
appearing in the thermal pressure, though not yet available.
The first few terms of this
expansion read\cite{Toimela:1985xy,Kap:FTFT}:
\begin{eqnarray}\label{cc0pt}
   {\chi \0 \chi_0}&=&1-{1\02}{3\0N}{N_g\08}\left(g\0\pi\right)^2+
{3\0N}{N_g\08}\sqrt{{N\03}+{N_f\0 6}}\left(g\0\pi\right)^3 \nn
&&-{3\04}{N_g\08}\left(g\0\pi\right)^4\log{1\0g}+ {\mathcal O}(g^4)
\end{eqnarray}
(for the more lengthy results on the subsequent coefficients we refer
the reader to Ref.~\cite{Vuorinen:2002ue}).

In these strictly
perturbative results the plasmon term overcompensates the
leading-order interaction term, much like in the case of the
pressure, though less severely so (for $N_f=2$ the diagonal QNS
fall below the ideal-gas value only when $T > 40 T_c$, and only
for $T>700 T_c$ does one find a growth of $\chi/\chi_0$
with temperature as observed on the lattice --- in the case
of the pressure these temperatures below which perturbation theory
clearly fails are orders of magnitudes higher).

The $\Phi$-derivable 2-loop expression (\ref{N2loop}) captures
both the leading-order term $\propto g^2$ and the plasmon term
$\propto g^3$ accurately, together with a subset of higher-order
terms that are usually expanded out in 2-loop perturbation theory.

As additional approximations we may start from the HTL quark
density $\mathcal N_{HTL}$ where (\ref{N2loop}) is evaluated
with HTL propagators. Differentiation with respect to the chemical
potential yields the corresponding approximation for the diagonal\footnote{In
the HTL approximation the QNS are purely diagonal.} QNS.

While this approximation contains the leading-order term $\propto g^2$
accurately, it does not include anything of the plasmon term.
In the QNS, the plasmon term arises completely from corrections
to the asymptotic thermal masses of quarks according to (\ref{dSigmaas}).

Numerically, the HTL approximation gives results somewhat above
the strictly perturbative result to order $g^2$, and, using
the same gap equations for the asymptotic thermal masses of
quarks as in the previous application, the result is halfway
towards the ideal-gas result, whereas a strictly perturbative
inclusion of the plasmon effect would have resulted in a result
in excess of the free one.

\begin{figure}[t]
\centerline{\includegraphics[bb=70 200 540 540,width=6.5cm]{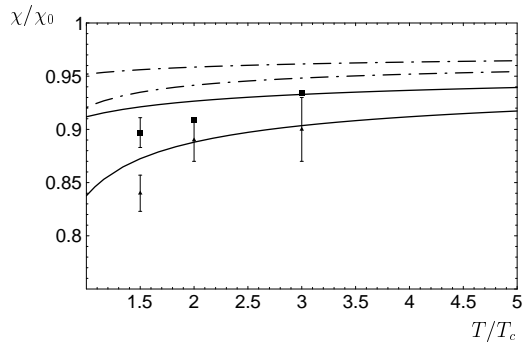}
}
\caption{Comparison of our results for $\chi/\chi_0$ 
in the formal limit $N_f$=$0$
with two recent continuum-extrapolated lattice results for 
quenced QCD\protect\cite{Gavai:2002jt}.
\label{fig:chinf0}}
\end{figure}

\begin{figure}
\centerline{\includegraphics[bb=70 200 540 540,width=6.5cm]{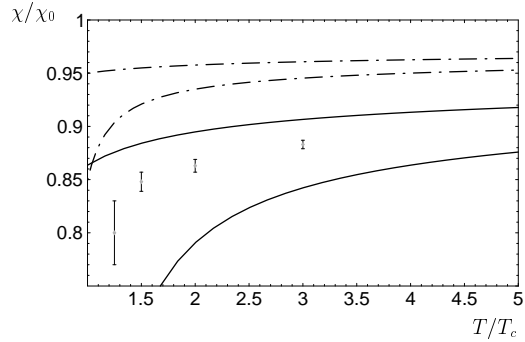}
}
\caption{
Comparison of our results for $\chi/\chi_0$ in massless $N_f$=2 QCD
with the lattice results of Ref.~\protect\cite{Gavai:2001ie}
(no continuum extrapolation).
Full lines refer to the HTL approximation, dash-dotted lines to
NLA, when the $\overline{\mbox{MS}}$ scale is varied from
$\bar\mu=\pi T$ to $\bar\mu=4\pi T$.
\label{fig:chinf2}}
\end{figure}

Fig.~\ref{fig:chinf0} shows the results for quenched QCD
together with recent results for two different continuum
extrapolations\cite{Gavai:2002jt} (both are higher than the previous lattice
results\cite{Gavai:2001fr}).
Fig.~\ref{fig:chinf2} shows our prediction\cite{Blaizot:2001vr}
for $N_f=2$ in comparison with the lattice data of Ref.~\cite{Gavai:2001ie},
for which a continuum extrapolation is not yet available, however.

It would be interesting to compare our predictions with HTLPT to two-loop
order; the one-loop HTL-resummed results from
charge correlators\cite{Chakraborty:2001kx
}
suffer from severe
overcounting while missing out the plasmon term\cite{Blaizot:2002xz}
so that one should not perform a comparison yet.%

\subsubsection{Off-diagonal quark number susceptibilities}

Since lattice results are starting to become available also for the
off-diagonal QNS, which in the limit of massless QCD we denote
collectively by $\tilde\chi$, we briefly discuss this quantity, too.
In contrast to the diagonal QNS, the ideal-gas value of
off-diagonal QNS is zero, so this directly probes the interactions
in the system.

The leading-order contribution to off-diagonal QNS
require two fermion loops connected by
gluon lines.
The diagram
with just one gluon exchange vanishes by colour neutrality.
The one with two gluon exchange is non-zero,
but because the fermion loops are then even functions
of $\mu$, it contributes to $\tilde\chi$ only when $\mu\not=0$,
starting at order $g^3$, namely
\be
\chi_{ij}={g^4 (N^2-1)T\mu_i \mu_j \0 16 \pi^5 m_D} \qquad \mbox{for
$i\not=j$}\,.
\ee
This contribution is indeed contained in the next-to-leading approximation
discussed above, but only after the corrections to the asymptotic
thermal masses have been included.

\begin{figure}[t]
\bigskip
\centerline{\includegraphics[width=2cm]{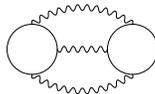}}
\caption{Lowest-order diagram in the thermodynamic potential
that contributes to off-diagonal susceptibilities
$\6{\mathcal N}_i/\6\mu_j$ at $\mu=0$.
\label{bugblatter}
}
\end{figure}

When all chemical potentials vanish (which is the case most easily
accessible to lattice studies),
the above approximations of the QNS become
strictly diagonal. In this limit
the lowest-order diagram contributing to off-diagonal QNS is
the 4-loop ``bugblatter'' diagram shown in Fig.~\ref{bugblatter}.%
\cite{Blaizot:2001vr} In this diagram each quark loop produces an effective
$C$-odd vertex for electrostatic gluons%
\cite{KorthalsAltes:1999cp
}
which in turn give rise to a logarithmically enhanced contribution
at order $g^6$, reading\cite{Blaizot:2001vr}
\be\label{tcc0}
{\tilde \chi\0\chi_0}\simeq-
{(N^2-1)(N^2-4)\0 128 N^2}\left({g\0\pi}\right)^6 \log{1\0g}.
\ee
This vanishes in SU(2) gauge theory
(though not in QED\cite{Blaizot:2001vr}), and in
(massless) QCD this leads to
\be\label{QCDtildechi}
{\tilde \chi\0\chi_0}\Big|_{N=3}
\simeq-{5\0144}\,\left({g\0\pi}\right)^6 \log{1\0g} 
\simeq-{10\09\pi^3}\,\alpha_s^3 \log(1/\alpha_s).
\ee

Assuming that the undetermined (though perturbatively computable)
constant under the log is of order one, the natural order of
magnitude of $\tilde\chi/\chi_0=\tilde \chi/T^2$ is $\sim 10^{-4}$
for $T\sim 3T_c$. Recent lattice studies in $N_f=2$ QCD%
\cite{Gavai:2001ie,Gavai:2002jt} have
found only values consistent with zero so far, with errors claimed to be
below $10^{-6}$, but this perhaps remains to be confirmed by independent
groups before it can be viewed as a new puzzle arising from
the comparison of lattice and perturbation theory.

\section{Conclusion and perspectives}

In this paper, we have reviewed progress made recently in the 
calculation of the high temperature
quark-gluon plasma using weak coupling techniques. Such calculations 
meet with difficulties, many
of which are not typical of gauge theories, but also occur in scalar 
field theories. In
fact, what determine the validity of a weak coupling expansion in 
thermal field theories is not only
the strength of the coupling, but also the magnitude of the thermal 
fluctuations which varies according to
the relevant momentum scales.

The thermodynamical functions are
dominated by hard degrees of freedom, carrying momenta of the order 
of the temperature $T$. Soft
modes, with typical momenta of order $gT$, contribute a correction, 
which however turns out to be
difficult to calculate, for essentially two reasons. First,  the 
dynamics of the soft
modes receive contributions from the hard modes 
which can only be taken into
account through all order resummations. Second, although the 
interactions among the soft modes
can be treated perturbatively, the corresponding  perturbation theory 
is an expansion in powers of
$g$ rather than
$g^2$, as it is for the hard modes. The latter feature is what causes 
the apparent poor convergence
of perturbation theory which has been observed in  calculations up to 
order $g^5$, both in the
scalar case and in QCD. 

In QCD, because of the unscreened magnetic fluctuations, perturbation 
theory eventually
breaks down. But this occurs at order $g^6$ only, and accordingly one 
could expect the contribution
of the  magnetic fluctuations to the pressure to be numerically 
small. This is in fact suggested
by detailed calculations based on dimensional reduction\cite{Kajantie:2002wa}. 
Therefore most of this review has
concentrated on the difficulties which are common to all thermal 
field theories, and  which involve the
momentum scale $k\sim gT$.

The
various resummation techniques which have been proposed to overcome 
the difficulties of perturbation theory
rely to some degree on a separation of scale between hard and soft excitations,
and follow the general pattern of effective theories. This is 
obviously the case for  the approaches based
on dimensional reduction, but is also true of screened perturbation 
theory and $\Phi$-derivable
approximations, although somewhat more indirectly. In  dealing with the soft
sector, most resummations use  the hard thermal loops as common 
building blocks.  Non-linear ``gap
equations'' for the calculation of  ``thermal mass'' parameters 
emerge in various forms, and the
difficulties associated with the ultraviolet renormalizations are 
dealt with in different fashions, depending
on the specific approximations.

In ``Screened Perturbation Theory'', and its extension to QCD 
referred to as ``HTL--perturbation
theory'', the use of the HTL  Lagrangian as a basis for the resummation
has the advantage to preserve manifest gauge symmetry, but also the 
drawback to lead to rather
complex  calculations beyond the lowest order. Besides the HTLs are 
used for {\it all}
momenta, soft and hard, while they are accurate only at soft momenta. 
At any finite order in this scheme,
one meets with temperature--dependent UV divergences whose renormalization
requires the use of ``thermal counterterms''. Although they  get in 
principle  corrected gradually as one
includes higher and higher orders,  these features remain 
unsatisfactory in the low order calculations that
one is able to perform.

At least at the level of the formalism, $\Phi$--derivable approximations
provides a conceptually clearer perspective on the general problem of 
propagator
renormalization than the simple ``add and subtract" approach of SPT. 
The overcounting is
automatically avoided by using skeleton expansions, and the gap 
equations emerge naturally, via the
variational principle for the thermodynamic potential. The HTLs 
appear as lowest--order
approximations to the solutions of the gap equations, and are used 
only in the kinematical domains
where they are legitimate approximations. However, the practical 
implementation of $\Phi$ derivable
approximations is not exempt of difficulties: In general, the gap 
equations  are {\it non--local}, which
complicates both their UV renormalization, and their numerical 
resolution. A further complication comes from
the fact that the
$\Phi$--derivable approximations violate  gauge symmetry. (However, 
this violation is parametrically
under control, which leaves the possibility to construct {\it 
approximate} solutions to the gap
equations which are gauge--invariant.)

One of the  most obvious advantages of  $\Phi$--derivable 
approximations remains the simplifications they
allow in the calculation of the entropy:  The two-loop calculation 
for the pressure turns into an
effectively one loop calculation for the entropy, which furthermore 
is finite once  the self-energy involved
in the calculation has been made finite.  The results
reported in Refs.\cite{Blaizot:1999ip,Blaizot:1999ap,Blaizot:2000fc}
have been obtained however after additional approximations. In the 
scalar 2--loop model,
which is exactly solvable, 
it has been possible to test these approximations by
comparing against the respective exact results. In QCD, where such a 
comparison is
not possible,  further refinements should be studied, in particular 
to deal with
non-local effects in the gap equations. Also, the calculation of the 
entropy is built on real-time
information, namely the dispersion relation for the hard 
quasiparticles. Understanding the relation with
purely Euclidean formalisms, such as that of dimensional reduction 
where this information does not enter in
an explicit fashion, is certainly worth further investigations.

In closing, one should perhaps emphasize that most of the 
calculations reported in this review point to the
validity of the {\it quasiparticle picture} of the quark-gluon plasma 
at high temperature. That is, they
support the idea that, by suitably dressing the elementary degrees of 
freedom, one could minimize the
effects of their residual interactions. This is particularly 
transparent in the entropy calculation. 
The results obtained in this manner for the thermodynamic potentials
and also for quark susceptibilities seem to encourage this physical picture,
which gives hope for the eventual applicability of correspondingly
improved perturbation theory to dynamical quantities.

{\small
\section*{Acknowledgments}

{\normalsize This work has been supported by the Austrian-French
scientific exchange program Amadeus of APAPE and \"OAD,
and the Austrian Science Foundation FWF, project no.~P14632.}
}



\begin{thebibliography}{100}
\newcommand{\enquote}[1]{``#1''}

\bibitem{Collins:1975ky}
J.~C. Collins and M.~J. Perry, {\it Phys. Rev. Lett.\/} {\bf 34}, 1353 (1975).

\bibitem{Karsch:2001cy}
F.~Karsch, {\it Lect. Notes Phys.\/} {\bf 583}, 209 (2002).

\bibitem{Laermann:2003cv}
E.~Laermann and O.~Philipsen, 
hep-ph/0303042.

\bibitem{Pisarski:2000eq}
R.~D. Pisarski, {\it Phys. Rev.\/} {\bf D62}, 111501 (2000);
P.~N. Meisinger, T.~R. Miller and M.~C. Ogilvie, {\it Phys. Rev.\/} {\bf D65},
  034009 (2002).

\bibitem{Kap:FTFT}
J.~I. Kapusta, {\it Finite-temperature field theory\/} (Cambridge University
  Press, Cambridge, UK, 1989).


\bibitem{LeB:TFT}
M.~{Le Bellac}, {\it Thermal Field Theory\/} (Cambridge University Press,
  Cambridge, UK, 1996).


\bibitem{Smilga:1996cm}
A.~V. Smilga, {\it Phys. Rept.\/} {\bf 291}, 1 (1997).

\bibitem{Blaizot:2001nr}
J.-P. Blaizot and E.~Iancu, {\it Phys. Rept.\/} {\bf 359}, 355 (2002).

\bibitem{Braaten:1990mz}
E.~Braaten and R.~D. Pisarski, {\it Nucl. Phys.\/} {\bf B337}, 569 (1990).

\bibitem{Frenkel:1990br}
J.~Frenkel and J.~C. Taylor, {\it Nucl. Phys.\/} {\bf B334}, 199 (1990).

\bibitem{Blaizot:1995ws}
J.-P. Blaizot, J.-Y. Ollitrault and E.~Iancu, in \enquote{Quark-gluon plasma},
  ed. R.~Hwa, vol.~2, pp. 135--210 (World Scientific, Singapore, 1995).


\bibitem{Blaizot:1993gn}
J.-P. Blaizot and E.~Iancu, {\it Nucl. Phys.\/} {\bf B390}, 589 (1993);
{\it Phys. Rev. Lett.\/} {\bf 70}, 3376 (1993).

\bibitem{Blaizot:1994be}
J.-P. Blaizot and E.~Iancu, 
{\it Nucl. Phys.\/} {\bf B417}, 608 (1994).

\bibitem{Nair:1993rx}
V.~P. Nair, {\it Phys. Rev.\/} {\bf D48}, 3432 (1993);
{\it Phys. Rev.\/} {\bf D50}, 4201 (1994).

\bibitem{Bodeker:1998hm}
D.~B{\"o}deker, {\it Phys. Lett.\/} {\bf B426}, 351 (1998);
{\it Nucl. Phys.\/} {\bf B559}, 502 (1999).

\bibitem{Blaizot:1999xk}
J.-P. Blaizot and E.~Iancu, {\it Nucl. Phys.\/} {\bf B557}, 183 (1999).

\bibitem{Kajantie:2000iz}
K.~Kajantie, M.~Laine, K.~Rummukainen and Y.~Schr{\"o}der, {\it Phys. Rev.
  Lett.\/} {\bf 86}, 10 (2001).

\bibitem{Karsch:1997gj}
F.~Karsch, A.~Patk{\'o}s and P.~Petreczky, {\it Phys. Lett.\/} {\bf B401}, 69
  (1997).

\bibitem{Chiku:1998kd}
S.~Chiku and T.~Hatsuda, {\it Phys. Rev.\/} {\bf D58}, 076001 (1998).

\bibitem{Andersen:2000yj}
J.~O. Andersen, E.~Braaten and M.~Strickland, {\it Phys. Rev.\/} {\bf D63},
  105008 (2001).

\bibitem{Andersen:1999fw}
J.~O. Andersen, E.~Braaten and M.~Strickland, {\it Phys. Rev. Lett.\/} {\bf
  83}, 2139 (1999);
%
{\it Phys. Rev.\/} {\bf D61},
  014017, 074016 (2000).

\bibitem{Baym:1962}
G.~Baym, {\it Phys. Rev.\/} {\bf 127}, 1391 (1962).


\bibitem{Blaizot:1999ip}
J.-P. Blaizot, E.~Iancu and A.~Rebhan, {\it Phys. Rev. Lett.\/} {\bf 83}, 2906
  (1999).

\bibitem{Blaizot:1999ap}
J.-P. Blaizot, E.~Iancu and A.~Rebhan, 
{\it Phys. Lett.\/} {\bf B470}, 181
  (1999).

\bibitem{Blaizot:2000fc}
J.-P. Blaizot, E.~Iancu and A.~Rebhan, {\it Phys. Rev.\/} {\bf D63}, 065003
  (2001).

\bibitem{Vanderheyden:1998ph}
B.~Vanderheyden and G.~Baym, {\it J. Stat. Phys.\/} {\bf 93}, 843 (1998).

\bibitem{Peshier:1996ty}
A.~Peshier, B.~K{\"a}mpfer, O.~P. Pavlenko and G.~Soff, {\it Phys. Rev.\/} {\bf
  D54}, 2399 (1996).

\bibitem{Levai:1997yx}
P.~Levai and U.~Heinz, {\it Phys. Rev.\/} {\bf C57}, 1879 (1998).

\bibitem{Peshier:1999ww}
A.~Peshier, B.~K{\"a}mpfer and G.~Soff, {\it Phys. Rev.\/} {\bf C61}, 045203
  (2000).

\bibitem{Schneider:2001nf}
R.~A. Schneider and W.~Weise, {\it Phys. Rev.\/} {\bf C64}, 055201 (2001).

\bibitem{Kastening:1997rg}
B.~Kastening, {\it Phys. Rev.\/} {\bf D56}, 8107 (1997).

\bibitem{Hatsuda:1997wf}
T.~Hatsuda, {\it Phys. Rev.\/} {\bf D56}, 8111 (1997).

\bibitem{Cvetic:2002ju}
G.~Cveti{\v c} and R.~K{\"o}gerler, {\it Phys. Rev.\/} {\bf D66}, 105009
  (2002).

\bibitem{Parwani:2000rr}
R.~R. Parwani, {\it Phys. Rev.\/} {\bf D63}, 054014 (2001);
{\it ibid.},
{\bf D64}, 025002 (2001).

\bibitem{Arnold:1994ps}
P.~Arnold and C.-X. Zhai, {\it Phys. Rev.\/} {\bf D50}, 7603 (1994).

\bibitem{Parwani:1995zz}
R.~Parwani and H.~Singh, {\it Phys. Rev.\/} {\bf D51}, 4518 (1995).

\bibitem{Braaten:1995cm}
E.~Braaten and A.~Nieto, {\it Phys. Rev.\/} {\bf D51}, 6990 (1995).

\bibitem{Kobes:1990xf}
R.~Kobes, G.~Kunstatter and A.~Rebhan, {\it Phys. Rev. Lett.\/} {\bf 64}, 2992
  (1990);
{\it Nucl. Phys.\/} {\bf B355}, 1
  (1991);
A.~Rebhan, {\it Lect. Notes Phys.\/} {\bf 583}, 161 (2002).

\bibitem{Rebhan:1993az}
A.~K. Rebhan, {\it Phys. Rev.\/} {\bf D48}, 3967 (1993).
{\it Nucl. Phys.\/} {\bf B430}, 319 (1994).

\bibitem{Blaizot:1995kg}
J.-P. Blaizot, E.~Iancu and R.~R. Parwani, {\it Phys. Rev.\/} {\bf D52}, 2543
  (1995).

\bibitem{Arnold:1995eb}
P.~Arnold and C.-X. Zhai, {\it Phys. Rev.\/} {\bf D51}, 1906 (1995).

\bibitem{Zhai:1995ac}
C.-X. Zhai and B.~Kastening, {\it Phys. Rev.\/} {\bf D52}, 7232 (1995).

\bibitem{Braaten:1996jr}
E.~Braaten and A.~Nieto, {\it Phys. Rev.\/} {\bf D53}, 3421 (1996).

\bibitem{Ginsparg:1980ef}
P.~Ginsparg, {\it Nucl. Phys.\/} {\bf B170}, 388 (1980).

\bibitem{Appelquist:1981vg}
T.~Appelquist and R.~D. Pisarski, {\it Phys. Rev.\/} {\bf D23}, 2305 (1981).

\bibitem{Nadkarni:1983kb}
S.~Nadkarni, {\it Phys. Rev.\/} {\bf D27}, 917 (1983).

\bibitem{Nadkarni:1988fh}
S.~Nadkarni, {\it Phys. Rev.\/} {\bf D38}, 3287 (1988).

\bibitem{Landsman:1989be}
N.~P. Landsman, {\it Nucl. Phys.\/} {\bf B322}, 498 (1989).

\bibitem{Braaten:1995na}
E.~Braaten, {\it Phys. Rev. Lett.\/} {\bf 74}, 2164 (1995).

\bibitem{Kajantie:1996dw}
K.~Kajantie, M.~Laine, K.~Rummukainen and M.~Shaposhnikov, {\it Nucl. Phys.\/}
  {\bf B458}, 90 (1996).

\bibitem{Rebhan:2000uc}
A.~Rebhan, in \enquote{Strong and Electroweak Matter 2000}, ed. C.~P.
  Korthals~Altes, pp. 199--203 (World Scientific, Singapore, 2000).

\bibitem{Luttinger:1960}
J.~M. Luttinger and J.~C. Ward, {\it Phys. Rev.\/} {\bf 118}, 1417 (1960).


\bibitem{DeDominicis:1964}
C.~{De Dominicis} and P.~C. Martin, {\it J. Math. Phys.\/} {\bf 5}, 14, 31
  (1964).


\bibitem{Cornwall:1974vz}
J.~M. Cornwall, R.~Jackiw and E.~Tomboulis, {\it Phys. Rev.\/} {\bf D10}, 2428
  (1974).

\bibitem{Braaten:2001en}
E.~Braaten and E.~Petitgirard, {\it Phys. Rev.\/} {\bf D65}, 041701, 085039 (2002).

\bibitem{vanHees:2001ik}
H.~van Hees and J.~Knoll, {\it Phys. Rev.\/} {\bf D65}, 025010, 105005 (2002);
{\it Phys. Rev.\/} {\bf D66}, 025028 (2002).

\bibitem{Blaizot:2003br}
J.-P. Blaizot, E.~Iancu and U.~Reinosa, 
hep-ph/0301201.

\bibitem{Drummond:1997cw}
I.~T. Drummond, R.~R. Horgan, P.~V. Landshoff and A.~Rebhan, {\it Nucl.
  Phys.\/} {\bf B524}, 579 (1998).

\bibitem{Andersen:2002ey}
J.~O. Andersen, E.~Braaten, E.~Petitgirard and M.~Strickland, {\it Phys.
  Rev.\/} {\bf D66}, 085016 (2002).

\bibitem{Andersen:2003zk}
J.~O. Andersen, E.~Petitgirard and M.~Strickland, 
hep-ph/0302069.

\bibitem{Riedel:1968}
E.~Riedel, {\it Z. Phys.\/} {\bf 210}, 403 (1968).


\bibitem{Kraemmer:1995az}
U.~Kraemmer, A.~K. Rebhan and H.~Schulz, {\it Ann. Phys.\/} {\bf 238}, 286
  (1995).

\bibitem{Kajantie:2002wa}
K.~Kajantie, M.~Laine, K.~Rummukainen and Y.~Schr{\"o}der, 
hep-ph/0211321.

\bibitem{vanRitbergen:1997va}
T.~van Ritbergen, J.~A.~M. Vermaseren and S.~A. Larin, {\it Phys. Lett.\/} {\bf
  B400}, 379 (1997).

\bibitem{Boyd:1996bx}
G.~Boyd {\it et~al.\/}, {\it Nucl. Phys.\/} {\bf B469}, 419 (1996).

\bibitem{Okamoto:1999hi}
M.~Okamoto {\it et~al.\/}, {\it Phys. Rev.\/} {\bf D60}, 094510 (1999).

\bibitem{Papa:1996an}
A.~Papa, {\it Nucl. Phys.\/} {\bf B478}, 335 (1996);
B.~Beinlich, F.~Karsch, E.~Laermann and A.~Peikert, {\it Eur. Phys. J.\/} {\bf
  C6}, 133 (1999).

\bibitem{Karsch:2000ps}
F.~Karsch, E.~Laermann and A.~Peikert, {\it Phys. Lett.\/} {\bf B478}, 447
  (2000).

\bibitem{AliKhan:2001ek}
A.~Ali~Khan {\it et~al.\/}, {\it Phys. Rev.\/} {\bf D64}, 074510 (2001).

\bibitem{Kajantie:1997tt}
K.~Kajantie, M.~Laine, K.~Rummukainen and M.~E. Shaposhnikov, {\it Nucl.
  Phys.\/} {\bf B503}, 357 (1997).

\bibitem{Polyakov:1978vu}
A.~M. Polyakov, {\it Phys. Lett.\/} {\bf B72}, 477 (1978).

\bibitem{Linde:1980ts}
A.~D. Linde, {\it Phys. Lett.\/} {\bf B96}, 289 (1980).

\bibitem{Gross:1981br}
D.~J. Gross, R.~D. Pisarski and L.~G. Yaffe, {\it Rev. Mod. Phys.\/} {\bf 53},
  43 (1981).

\bibitem{Blaizot:2003iq}
J.-P. Blaizot, E.~Iancu and A.~Rebhan, 
hep-ph/0303045.

\bibitem{Laine:2003ay}
M.~Laine, 
hep-ph/0301011.

\bibitem{Braaten:1992gm}
E.~Braaten and R.~D. Pisarski, {\it Phys. Rev.\/} {\bf D45}, 1827 (1992).

\bibitem{Frenkel:1992ts}
J.~Frenkel and J.~C. Taylor, {\it Nucl. Phys.\/} {\bf B374}, 156 (1992).

\bibitem{Baier:1999db}
R.~Baier and K.~Redlich, {\it Phys. Rev. Lett.\/} {\bf 84}, 2100 (2000).

\bibitem{Andersen:2001ez}
J.~O. Andersen and M.~Strickland, {\it Phys. Rev.\/} {\bf D64}, 105012 (2001).

\bibitem{Freedman:1977xs}
B.~A. Freedman and L.~D. McLerran, {\it Phys. Rev.\/} {\bf D16}, 1130 (1977).

\bibitem{Arrizabalaga:2002hn}
A.~Arrizabalaga and J.~Smit, {\it Phys. Rev.\/} {\bf D66}, 065014 (2002).

\bibitem{Peshier:2000hx}
A.~Peshier, {\it Phys. Rev.\/} {\bf D63}, 105004 (2001).

\bibitem{Kalashnikov:1980cy}
O.~K. Kalashnikov and V.~V. Klimov, {\it Sov. J. Nucl. Phys.\/} {\bf 31}, 699
  (1980).

\bibitem{Weldon:1982aq}
H.~A. Weldon, {\it Phys. Rev.\/} {\bf D26}, 1394 (1982).

\bibitem{Pisarski:1989cs}
R.~D. Pisarski, {\it Physica\/} {\bf A158}, 146 (1989).

\bibitem{Klimov:1981ka}
V.~V. Klimov, {\it Sov. J. Nucl. Phys.\/} {\bf 33}, 934 (1981).

\bibitem{Pisarski:1989wb}
R.~D. Pisarski, {\it Nucl. Phys.\/} {\bf A498}, 423C (1989).

\bibitem{Blaizot:1993bb}
J.-P. Blaizot and J.-Y. Ollitrault, {\it Phys. Rev.\/} {\bf D48}, 1390 (1993).

\bibitem{Kraemmer:1990dr}
U.~Kraemmer, M.~Kreuzer and A.~Rebhan, {\it Ann. Phys.\/} {\bf 201}, 223
  (1990).

\bibitem{Flechsig:1996ju}
F.~Flechsig and A.~K. Rebhan, {\it Nucl. Phys.\/} {\bf B464}, 279 (1996).

\bibitem{Schulz:1994gf}
H.~Schulz, {\it Nucl. Phys.\/} {\bf B413}, 353 (1994).

\bibitem{Moore:2002md}
G.~D. Moore, {\it JHEP\/} {\bf 0210}, 055 (2002);
A.~Ipp, G.~D. Moore and A.~Rebhan, {\it JHEP\/} {\bf 0301}, 037 (2003).

\bibitem{Rebhan:2003fj}
A.~Rebhan, 
hep-ph/0301130.

\bibitem{Karsch:1999vy}
F.~Karsch, {\it Nucl. Phys. Proc. Suppl.\/} {\bf 83}, 14 (2000).

\bibitem{Drummond:1999si}
I.~T. Drummond, R.~R. Horgan, P.~V. Landshoff and A.~Rebhan, {\it Phys.
  Lett.\/} {\bf B460}, 197 (1999).

\bibitem{Fodor:2001au}
Z.~Fodor and S.~D. Katz, {\it Phys. Lett.\/} {\bf B534}, 87 (2002);
{\it JHEP\/} {\bf 0203}, 014 (2002);
P.~de~Forcrand and O.~Philipsen, {\it Nucl. Phys.\/} {\bf B642}, 290 (2002);
C.~R. Allton {\it et~al.\/}, {\it Phys. Rev.\/} {\bf D66}, 074507 (2002).

\bibitem{Fodor:2002km}
Z.~Fodor, S.~D. Katz and K.~K. Szabo, 
hep-lat/0208078.

\bibitem{Szabo:2003kg}
K.~K. Szabo and A.~I. Toth, 
  hep-ph/0302255.

\bibitem{Romatschke:2002pb}
P.~Romatschke, 
hep-ph/0210331.

\bibitem{Asakawa:2000wh}
M.~Asakawa, U.~W. Heinz and B.~M{\"u}ller, {\it Phys. Rev. Lett.\/} {\bf 85},
  2072 (2000);
S.~Jeon and V.~Koch, {\it Phys. Rev. Lett.\/} {\bf 85}, 2076 (2000);
M.~Prakash, R.~Rapp, J.~Wambach and I.~Zahed, {\it Phys. Rev.\/} {\bf C65},
  034906 (2002).

\bibitem{Gavai:2001ie}
R.~V. Gavai, S.~Gupta and P.~Majumdar, {\it Phys. Rev.\/} {\bf D65}, 054506
  (2002).

\bibitem{Gavai:2002kq}
R.~V. Gavai and S.~Gupta, {\it Phys. Rev.\/} {\bf D65}, 094515 (2002).

\bibitem{Gavai:2002jt}
R.~V. Gavai and S.~Gupta, 
hep-lat/0211015.

\bibitem{Bernard:2002yd}
C.~Bernard {\it et~al.\/}, 
hep-lat/0209079.

\bibitem{Vuorinen:2002ue}
A.~Vuorinen, 
hep-ph/0212283.

\bibitem{Toimela:1985xy}
T.~Toimela, {\it Int. J. Theor. Phys.\/} {\bf 24}, 901 (1985).

\bibitem{Blaizot:2001vr}
J.-P. Blaizot, E.~Iancu and A.~Rebhan, {\it Phys. Lett.\/} {\bf B523}, 143
  (2001).

\bibitem{Gavai:2001fr}
R.~V. Gavai and S.~Gupta, {\it Phys. Rev.\/} {\bf D64}, 074506 (2001).

\bibitem{Chakraborty:2001kx}
P.~Chakraborty, M.~G. Mustafa and M.~H. Thoma, {\it Eur. Phys. J.\/} {\bf C23},
  591 (2002);
  hep-ph/0303009.

\bibitem{Blaizot:2002xz}
J.-P. Blaizot, E.~Iancu and A.~Rebhan, 
  hep-ph/0206280.

\bibitem{KorthalsAltes:1999cp}
C.~P. Korthals-Altes, R.~D. Pisarski and A.~Sinkovics, {\it Phys. Rev.\/} {\bf
  D61}, 056007 (2000);
A.~Hart, M.~Laine and O.~Philipsen, {\it Nucl. Phys.\/} {\bf B586}, 443 (2000);
D.~B{\"o}deker and M.~Laine, {\it JHEP\/} {\bf 0109}, 029 (2001).

\end{thebibliography}
\end{document}